\documentclass[12pt]{iopart}

\usepackage{graphicx}
\usepackage{cite}
\usepackage{color}
\usepackage{multirow}
\usepackage{amssymb}
\pdfoutput=1  
\begin{document}
\bibliographystyle{iopart-num}

\title[]{The effect of the driving frequency on the confinement of beam electrons and plasma density in low pressure capacitive discharges}

\author{S. Wilczek$^1$, J. Trieschmann$^1$, J. Schulze$^2$, E. Schuengel$^2$, R. P. Brinkmann$^1$, A. Derzsi$^3$, I. Korolov$^3$, Z. Donk\'o$^3$, and T. Mussenbrock$^1$}
\address{$^1$Institute for Theoretical Electrical Engineering, Ruhr-University Bochum, Germany. \\ 
$^2$Department of Physics, West Virginia University, Morgantown, WV 26506, USA. \\ 
$^3$Institute for Solid State Physics and Optics, Wigner Research Centre for Physics,
Hungarian Academy of Sciences, 1121 Budapest, Konkoly Thege Mikl\'os str. 29-33, Hungary.}
\ead{Sebastian.Wilczek@rub.de}

\begin{abstract}
The effect of changing the driving frequency on the plasma density and the electron dynamics in a capacitive radio-frequency argon plasma operated at low pressures of a few Pa is investigated by Particle-in-Cell/Monte-Carlo Collision simulations and analytical modeling. In contrast to previous assumptions, the plasma density does not follow a quadratic dependence on the driving frequency in this non-local collisionless regime. Instead, a step-like increase at a distinct driving frequency is observed. Based on an analytical power balance model, in combination with a detailed analysis of the electron kinetics, the density jump is found to be caused by an electron heating mode transition from the classical $\alpha$-mode into a low density resonant heating mode characterized by the generation of two energetic electron beams at each electrode per sheath expansion phase. These electron beams propagate through the bulk without collisions and interact with the opposing sheath. In the low density mode, the second beam is found to hit the opposing sheath during its collapse. Consequently, a high number of energetic electrons is lost at the electrodes resulting in a poor confinement of beam electrons in contrast to the classical $\alpha$-mode observed at higher driving frequencies. Based on the analytical model this modulated confinement quality and the related modulation of the energy lost per electron lost at the electrodes is demonstrated to cause the step-like change of the plasma density. The effects of a variation of the electrode gap, the neutral gas pressure, the electron sticking and secondary electron emission coefficients of the electrodes on this step-like increase of the plasma density are analyzed based on the simulation results.  
\end{abstract}

\submitto{Plasma Sources Science and Technology}

\maketitle

\section{Introduction}
Capacitively coupled radio frequency (CCRF) discharges operated at low pressures are highly relevant for industrial applications such as plasma etching and sputtering\cite{LiebermanBook,ChabertBook,MakabeBook,RFSputter}.
For these purposes low neutral gas pressures of a few Pascal are typically used to ensure a perpendicular energetic ion bombardment of the wafer. Commercial reactors are designed to optimize process rates via providing a high plasma density and ion flux.
In this context, the effects of changing the driving frequency, $f$, and the electrode gap, $d$, on the plasma density in this low pressure non-local and collisionless regime are highly relevant for fundamental insights into the physics of CCRF discharges and applications, e.g., for improved reactor designs. Changing these two global control parameters affects the electron heating dynamics and the plasma density. At such low pressure conditions the electron heating is typically dominated by stochastic \cite{HWModel2,HWModel,PrHeat1,PrHeat2,PrHeat3,Kaganovich1,Kaganovich2,Mussenbrock,Turner1,Turner2,Lafleur1,Lafleur2,Lafleur3,Schulze1,Schulze2} and ambipolar heating \cite{AmbHeat} during the phases of sheath expansion at both electrodes in electropositive discharges. This heating mode is typically called the $\alpha$-mode \cite{Belenguer1990}. In case of high ion fluxes and limited electron mobilities electric field reversals during sheath collapse can also cause significant electron heating \cite{LiebermanFieldRev,VenderFieldRev,TurnerFieldRev,FieldReversalSchulze,UCZFieldRev,GrahamFieldRev}. Changing global control parameters such as the neutral gas pressure, the frequency, or the gas mixture can induce transitions into different heating modes. These can occur as the $\gamma$-mode at high pressures, low driving frequencies and/or high driving voltage amplitudes\cite{Belenguer1990,Schulze_Gamma,EAEF,Voloshin1} or the Drift-Ambipolar heating mode at high pressures and/or in electronegative plasmas\cite{Schulze_Omega,Hemke_Omega,Boeuf_Omega,Voloshin2}.

In this work we investigate a regime, where stochastic electron heating dominates. Extensive research on this topic has been performed in the past: by experiments and models, Lieberman and Godyak have shown that electrons are heated collisionlessly on time average by repeated interactions with the electric field of the time modulated plasma sheaths \cite{HWModel2}. Based on a global power balance, the plasma density is assumed to increase proportionally to $f^{2}$ at fixed pressure and electrode gap\cite{LiebermanBook,ChabertBook}. Such models include the energy loss per electron lost at the electrodes, $\varepsilon_{\rm{e}}$, as one loss term and assume it to be dominated by thermal electrons. Under the assumption of a Maxwellian Electron Energy Distribution Function (EEDF) $\varepsilon_{\rm{e}} \approx 2 k_B T_{\rm{e}}$ is assumed, where $T_{\rm{e}}$ is the electron temperature. Under the low pressure conditions investigated here the EEDF is non-Maxwellian and the mean electron energy is strongly space dependent and changes on a nanosecond timescale due to the presence of non-local kinetic effects so that the definition of a global electron temperature becomes meaningless (Correspondingly, the terms heating and cooling are used with respect to the total kinetic energy of the electrons, not the electron temperature). In fact, beams of highly energetic electrons are generated at each electrode during sheath expansion and propagate almost collisionlessly through the bulk until they hit the opposing sheath, since $\lambda_{\rm{m}}/d > 1$ \cite{WoodBeams,FTCBeams,IEEEBeams}, where $\lambda_{\rm{m}}$ is the electron mean free path. Such kinetic effects are neglected in previous models, but are found to play a critical role when the gas pressure is low.

The impingement phase of these highly energetic electrons at the opposing sheath is found to be of great relevance. If an energetic beam electron hits the opposing sheath at a phase of high local sheath potential, it will be reflected back into the bulk and can still ionize the neutral gas, i.e. its confinement is effective. While interacting with the moving sheath, it can be cooled (collapsing sheath) or heated (expanding sheath). If an energetic electron bounces back and forth between both sheaths and hits each sheath during its expansion phase, it will be heated multiple times. This effect is called electron Bounce Resonance Heating (BRH) and was recently investigated by Liu et al. \cite{BRH1,BRH2,BRH3,BRH4,BRH_Hysteresis} based on earlier work of Wood\cite{WoodBeams,Wood2}. They demonstrated that an electron beam with a certain velocity will be heated coherently between both sheath expansions, if the time to traverse the bulk, $\tau$, is half the RF period. It was also confirmed that electron BRH is affected by the electrode gap distance, the pressure and the driving frequency and affects the plasma density. If beam electrons hit the opposing sheath during sheath collapse, i.e., at a phase of low local sheath voltage, they will reach the electrode and will be lost from the plasma, i.e., their confinement is not effective. 

In this work, we study the effect of changing the driving frequency on the plasma density and the electron heating dynamics by Particle-In-Cell/Monte-Carlo Collision (PIC/MCC) simulations and an analytical power balance model, at low pressures in argon gas. In contrast to previous models, we find that the plasma density is not proportional to $f^{2}$ in this collisionless regime, where $\lambda_{\rm{m}}/d> 1$. Instead, a complex behavior of the plasma density as a function of the applied radio-frequency is observed including a steep step-like increase at a distinct driving frequency. Based on the power balance model we demonstrate that this step-like increase is caused by a modulation of the confinement quality of energetic beam electrons as a function of the driving frequency. An abrupt decrease of the plasma density is observed, when a high number of energetic beam electrons hits the opposing collapsed sheath and is lost at the electrodes. This causes the energy loss per electron lost at the electrodes, $\varepsilon_e$, to increase drastically and the plasma density to decrease strongly. The number of energetic electron beams generated at a given electrode during one phase of sheath expansion and the impingement phase of beam electrons at the opposing electrode are found to change as a function of the driving frequency, since an electron heating mode transition is induced by changing the frequency. At high driving frequencies and high plasma densities the sheaths are small and expand relatively slowly, while the sheaths are found to expand more rapidly at lower driving frequencies and low plasma densities under the conditions investigated here. Below a distinct driving frequency the sheaths expand so fast that the electrons cannot react and oscillations of the electron heating rate at frequencies comparable to the local electron plasma frequency are excited similarly to previous findings of Vender et al. \cite{VenderFieldRev,Vender2}. Based on Ku et al. \cite{Annaratone,Ku,Ku2} these oscillations might be considered to be consequences of the local excitation of a Plasma Parallel Resonance (PPR). This effect induces a transition into a resonant heating mode. In this low density resonant mode two electron beams are generated at each electrode during one phase of sheath expansion. The later beam hits the opposing electrode during sheath collapse and causes the observed increase of $\varepsilon_e$ and, thus, the decrease of the density. This mechanism is self-amplifying and keeps the discharge in the low density resonant heating mode. Such kinetic confinement effects are not included in previous models, however, they have a dramatic effect on the plasma density. We investigate the effects of changing the electrode gap, the neutral gas pressure, and the electron reflection as well as the ion induced secondary electron emission coefficients at both electrodes on the plasma density at different driving frequencies.  

The paper is structured in the following way: In section 2 the PIC/MCC simulation and the choice of input parameters is introduced briefly. The analytical power balance model is described in section 3. The results are presented and discussed in section 4. Finally, conclusions are drawn in section 5.\\


\section{PIC/MCC setup}
PIC/MCC simulations provide the basis for studying the kinetic behavior of electron beams in CCRF plasmas based on a self-consistent and accurate description of the particle dynamics. In low temperature plasmas kinetic and non-local effects are of great relevance. The trajectories of fast electrons and their interaction with the plasma sheaths, as well as the kinetic energy loss due to electrons lost at the walls can be obtained from these simulations.
In this work we use a 1d3v electrostatic Particle-In-Cell code, which has been benchmarked against other codes developed by different authors\cite{benchmark}. The code includes three types of electron-neutral (elastic, excitation and ionization) and two types of ion-neutral collisions (isotropic and backward elastic scattering) for an argon plasma. The required cross sections are taken from the JILA database\cite{Phelps,sim3}. All particle interactions are treated by the Monte-Carlo method with a null-collision scheme \cite{Mertmann,Birdsall,NC}. The electrodes are assumed to be infinite, planar and parallel. One of the electrodes is driven by the following voltage waveform:

\begin{equation}
\phi(t) = \phi_0 \cos{(2 \pi f t)},
\end{equation}

while the other electrode is grounded. 

The electrode gap, $d$, is varied (14 mm $< d <$ 20 mm) and the neutral gas pressure is between 1 Pa and 5 Pa.
The driving frequency, $f$, is scanned from 40 MHz to 90 MHz, while the applied voltage amplitude is kept constant at $\phi_0 = 150$ V. 
We define a base case with the following parameters: 1.3 Pa, 1.5 cm electrode gap, and 150 V voltage amplitude; the electron and ion sticking coefficients are set to 1 (no reflection) and the secondary electron emission from the electrodes is neglected due to the low pressure scenario investigated. By initially switching off these surface reactions in the simulation a clearer analysis of the underlying physics is possible due to the absence of surface induced effects. After understanding the fundamental particle dynamics of this base case, the electron sticking coefficient, $s$, and the ion induced secondary electron emission coefficient, $\gamma$, are varied and their effects on the plasma density at different driving frequencies are studied. $s$ is varied from $1$ to $0.2$ and $\gamma$ is varied from $0$ to $0.2$, to cover the range of surface coefficients typically used in simulations of CCRF plasmas \cite{Schulze_Gamma,Kollath_Alpha,Petrovic_Gamma} and also relevant to other physical systems. Identical electrode materials are assumed, i.e. the electrode surface coefficients are kept identical at both electrodes.
The spatial grid used in the simulations is $\Delta x = d/N_{cells}$, where $N_{cells} = 512$ is the number of numerical grid cells. We set $N_{\rm{tspc}}=$ 2000-4000 timesteps per RF period and the time step is $\Delta t = (f \cdot N_{\rm{tspc}})^{-1}$. 
As we vary the driving frequency and the gap size, the space and time step, $\Delta x$ and $\Delta t$, are not fixed, but fulfill the following stability and accuracy conditions, while minimizing the required computational time:
\begin{eqnarray}
\omega_{\rm{pe}}\Delta t &\lesssim& 0.2,\\
\frac{\Delta x}{\lambda_{\rm{D}}} &\lesssim& 0.5.
\end{eqnarray}
Here, $\lambda_{\rm{D}}$ is the Debye length and $\omega_{\rm{pe}}$ is the electron plasma frequency. We use an adaptive particle weighting in order to obtain more than 100000 super-particles (for electrons) when convergence is achieved.

\section{Electron Power Balance Model}

We use a power balance for the electrons to calculate the ion density at the electrodes, $n_{\rm{i,el}}$, as a function of the driving frequency, while keeping the driving voltage amplitude, the pressure, and the electrode gap constant\cite{LiebermanBook,ChabertBook}:

\begin{equation}
S_{\rm{abs}} = 2 e n_{\rm{i,el}} u_{\rm{i,el}} (\varepsilon_{\rm{c}} + \varepsilon_{\rm{e}})
\end{equation}

Here, $S_{\rm{abs}}$ is the total power absorbed by the electrons per area which is calculated by the space and time average product $\langle E\cdot J_{e} \rangle_{t,x}$, where $E$ is the electric field and $J_{e}$ is the electron current density. $e$ is the elementary charge, $u_{\rm{i,el}}$ is the ion velocity at the electrodes, $\varepsilon_{\rm{c}}$ is the collisional electron energy loss per electron-ion pair created, and $\varepsilon_{\rm{e}}$ is the average energy per electron lost at the electrodes. This model assumes equal electron and ion fluxes to each electrode on time average. This condition is necessarily fulfilled in  CCRF plasmas driven via a blocking capacitor \cite{CDyn}. Solving this equation for $n_{\rm{i,el}}$ allows to calculate the ion density at the electrodes: 

\begin{equation}
n_{\rm{i,el}} = \frac{S_{\rm{abs}}}{2 e u_{\rm{i,el}} (\varepsilon_{\rm{c}} + \varepsilon_{\rm{e}})}
\label{Model},
\end{equation}

where $S_{\rm{abs}}$, $u_{\rm{i,el}}$, $\varepsilon_{\rm{e}}$ and $\varepsilon_{\rm{c}}$ are taken as input parameters from the PIC/MCC simulation.

Reproducing the plasma density as a function of the driving frequency obtained by the simulation via this model allows to understand the physical origin of the sudden increase in density at distinct frequencies observed in the simulation. This is carried out by analyzing the absorbed power per area, the ion velocity at the electrodes, the energy per electron lost at the electrodes, and the collisional energy loss per electron-ion pair created separately as a function of the driving frequency. The model then allows to determine the effect of each of these parameters on the ion density at the electrodes.

\section{Results}

\subsection{The effect of energetic beam electron confinement on the plasma density}

\begin{figure}[h!]
\begin{center}
\begin{tabular}{cc}
  \includegraphics[width=0.425\textwidth]{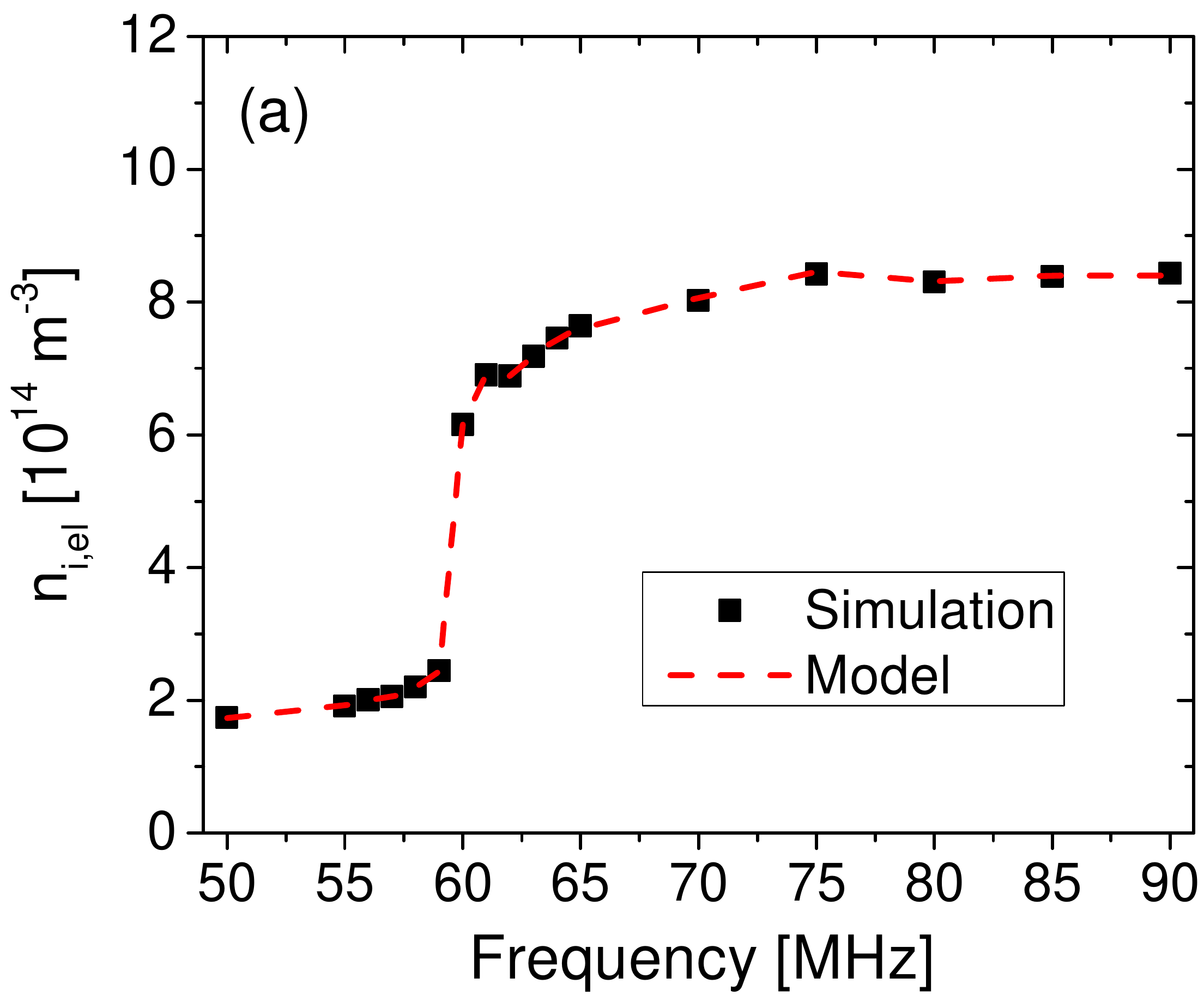}
 &
  \includegraphics[width=0.5\textwidth]{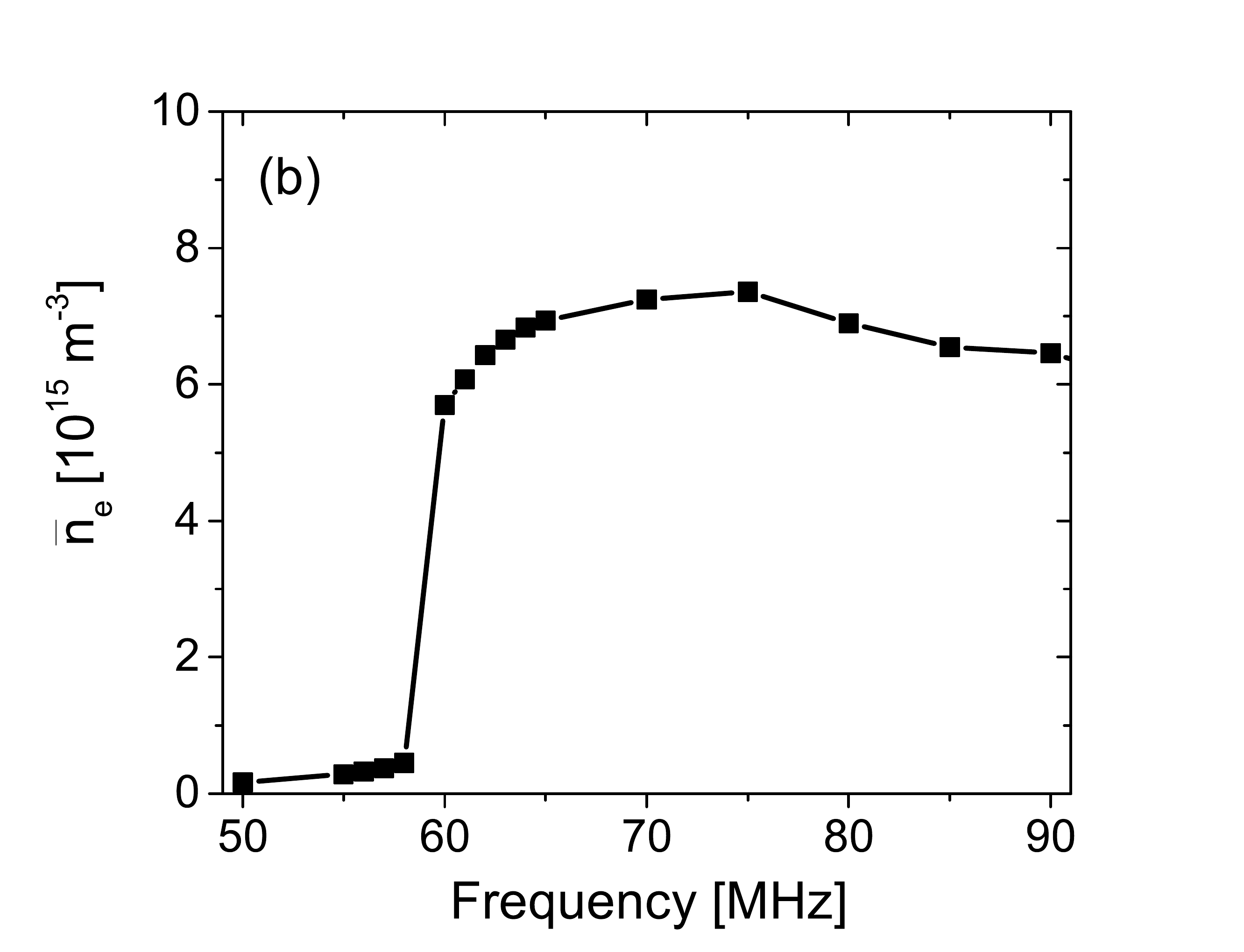}
 \\
\end{tabular}
\caption{Left: Ion density, $n_{i,el}$ at the electrodes as a function of the driving frequency obtained from the PIC/MCC simulation and the model. Right: Space and time averaged electron density, $\bar{n}_e$, as a function of the driving frequency obtained from the simulation. Conditions: $p = 1.3$ Pa, $d = 1.5$ cm, $\phi_{0} = 150$ V, $s = 1$, and $\gamma = 0$.}
\label{Density}
\end{center}
\end{figure}

The left plot of figure \ref{Density} shows the ion density at the electrodes obtained from the simulation and the model as a function of the driving frequency, at 1.3 Pa, 1.5 cm electrode gap, $\phi_{0} = 150$ V. Electron reflection and secondary electron emission at the electrodes are switched off in the simulation. The right plot of figure \ref{Density} shows the space and time averaged electron density, $\bar{n}_{\rm{e}} = (d_{gap} T_{\rm{RF}})^{-1} \int_0^{d_{gap}}{\int_0^{T_{\rm{RF}}}{n_{\rm{e}}(x,t) dt dx}}$, as a function of the driving frequency under the same conditions. Classical models predict a quadratic dependence of the plasma density on the driving frequency, which would correspond to a parabola in figure \ref{Density}\cite{LiebermanBook,ChabertBook}. Our results clearly demonstrate that this is not the case under the low pressure non-local conditions studied here. Instead, a step-like increase of the plasma density is observed around 60 MHz. Changing the driving frequency by 1 MHz from 59 MHz to 60 MHz leads to an increase of the ion density at the electrodes by a factor of about 4 and an increase of the electron density on space and time average by a factor of about 13. Such step-like increases of the plasma density have been previously observed experimentally and in PIC simulations as a function of the electrode gap under similar low pressure conditions in dual frequency discharges \cite{BRH1,BRH2,BRH3,BRH4,BRH_Hysteresis}. In the experiment a floating double probe was used for the measurement of the ion density and the light intensity was determined via optical emission spectroscopy. Based on the comparison of experiments and simulations, they determined a maximum in these quantities for a certain gap size, where the confinement of highly energetic electrons was most efficient (Bounce Resonance Effect). When they decreased the gap size from this value the ion density, as well as the light intensity decreased dramatically.  
Sudden changes of these characteristics are highly relevant for applications of CCRF discharges at low pressures such as plasma etching and RF sputtering, for which the plasma is operated in this highly non-local regime. In these applications typically a high ion flux and plasma density are required, while keeping the driving frequency low enough to avoid lateral inhomogeneities due to electromagnetic effects such as the Standing Wave Effect \cite{EM_Chabert,EM_Perret,EM_Eremin,EM_Lieberman,EM_Sansonnens}. Our results show that a small difference in the driving frequency can cause a great increase in plasma density and, therefore, potentially a significant increase in process rates and performance.     

\begin{figure}[h!]
\begin{center}
\begin{tabular}{cc}
  \includegraphics[width=0.5\textwidth]{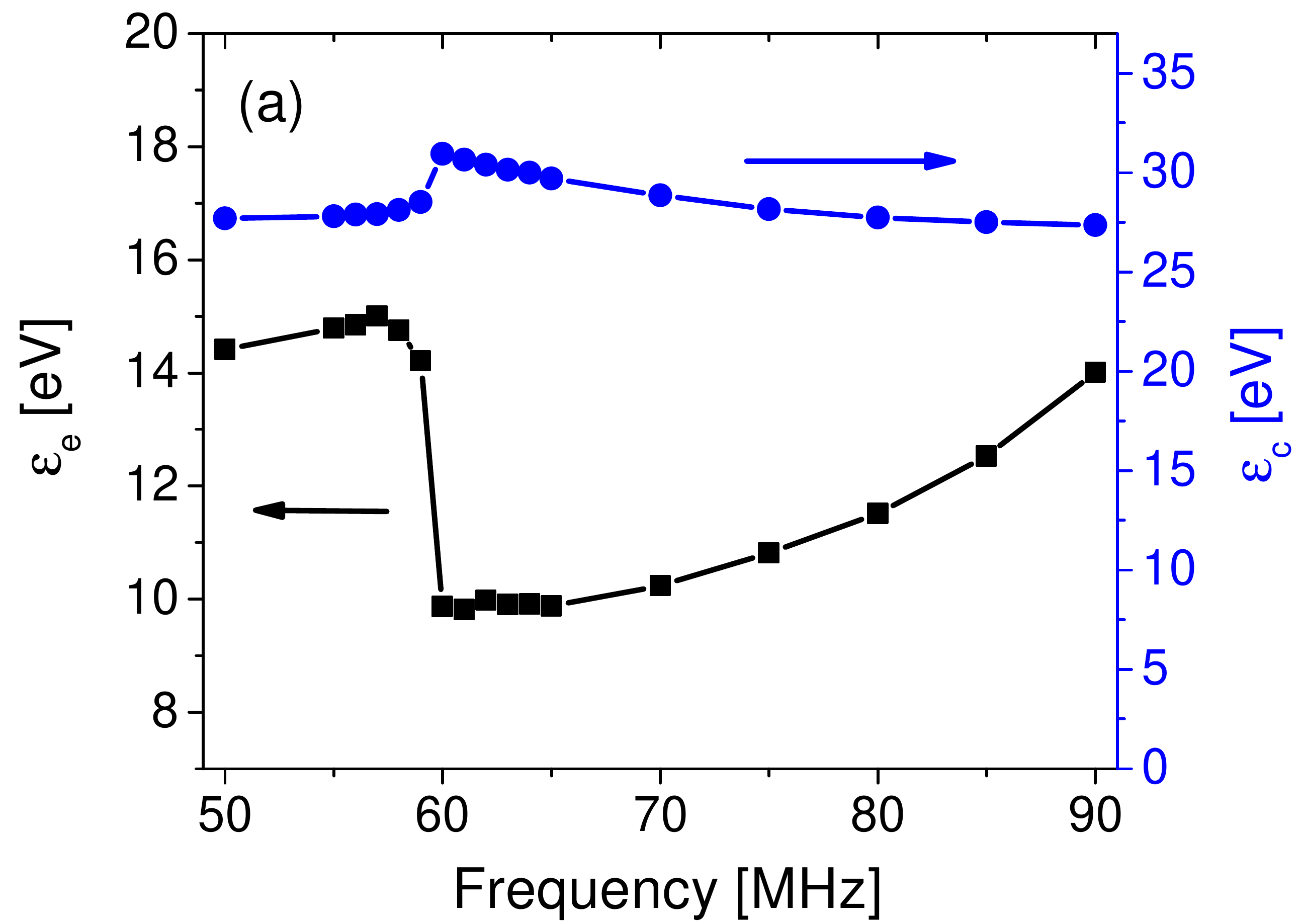}
 &
  \includegraphics[width=0.5\textwidth]{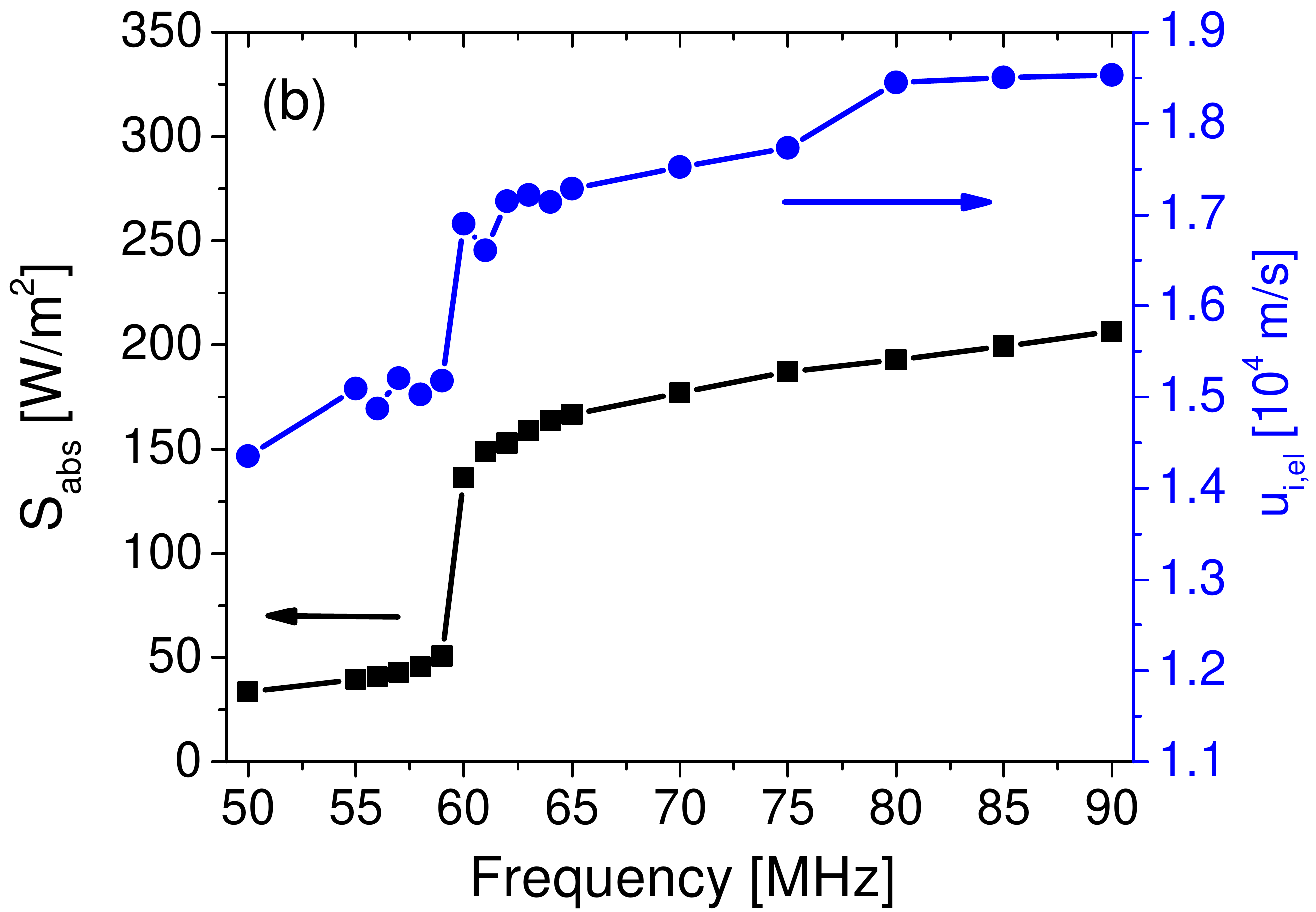}
 \\
\end{tabular}
\caption{Left: Energy per electron lost at the electrodes, $\varepsilon_{\rm{e}}$, and collisional energy loss per electron - ion pair, $\varepsilon_{\rm{c}}$, as a function of the driving frequency. Right: Power absorbed by the electrons per area, $S_{\rm{abs}}$, and ion velocity at the electrodes, $u_{\rm{i,el}}$, as a function of the driving frequency. All data are obtained from the simulation. Conditions: $p = 1.3$ Pa, $d = 1.5$ cm, $\phi_{0} = 150$ V, $s = 1$, and $\gamma = 0$.}
\label{ModelTerms}
\end{center}
\end{figure}

\begin{figure}[h!]
\begin{center}
\begin{tabular}{cc}
  \includegraphics[width=0.5\textwidth]{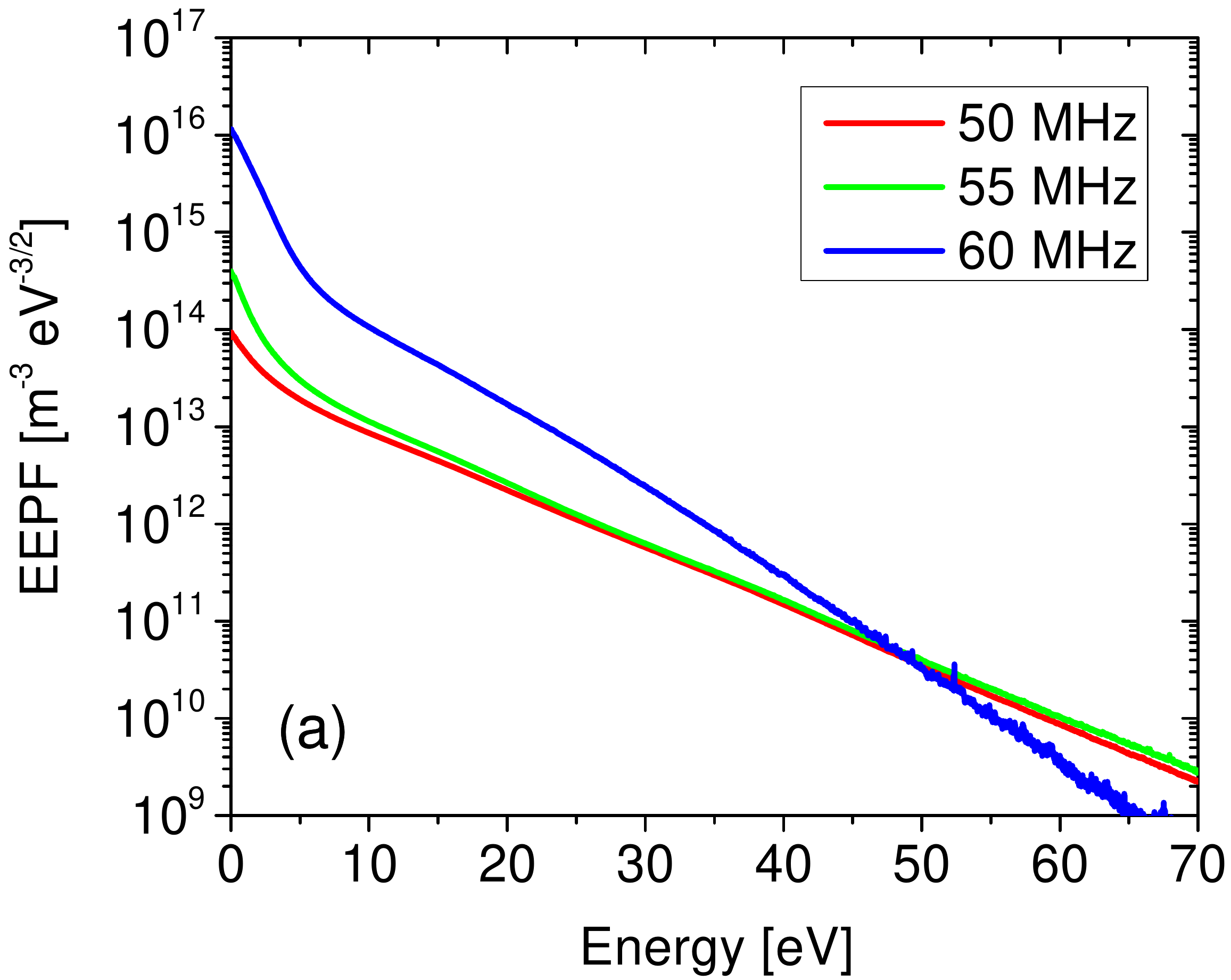}
 &
  \includegraphics[width=0.5\textwidth]{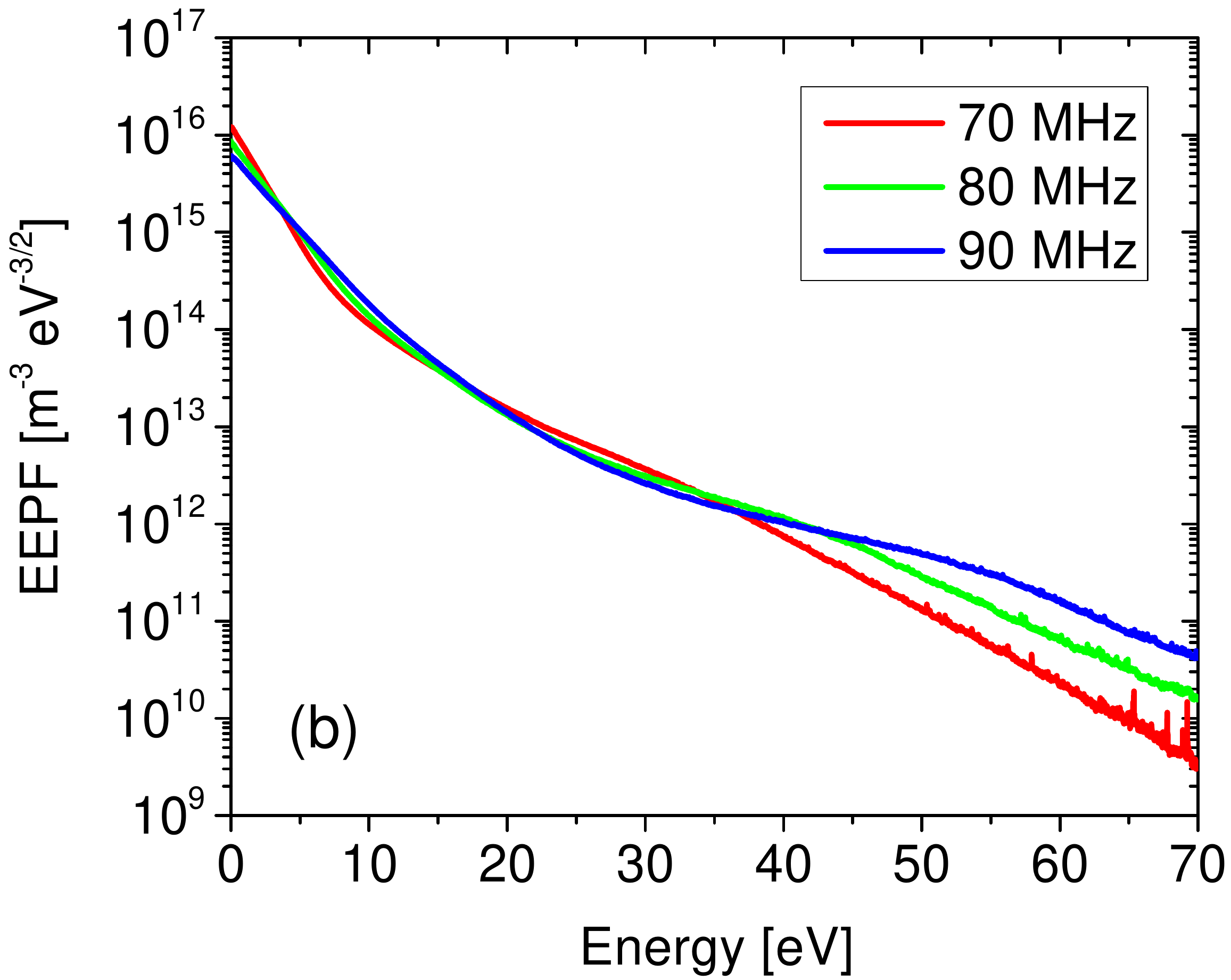}
 \\
\end{tabular}
\caption{Time averaged EEPF in the whole discharge volume for different driving frequencies obtained from the simulation. Conditions: $p = 1.3$ Pa, $d = 1.5$ cm, $\phi_{0} = 150$ V, $s = 1$, and $\gamma = 0$.}
\label{EEDF}
\end{center}
\end{figure}

Based on equation (\ref{Model}) the model reproduces the ion density at the electrodes obtained from the simulation perfectly and provides the basis for a detailed understanding of the physical origin of its step-like increase at $f = 60$ MHz. Figure \ref{ModelTerms} shows the input parameters of the model taken from the simulation as a function of the driving frequency. In combination with a more detailed analysis of the transient electron dynamics based on the simulation data, these diagrams reveal the physical origin of the density jump. The step-like increase of the plasma density is caused by two effects: (i) The energy loss per electron lost to the electrodes drops by about 5 eV ($\approx 30$ \%) near 60 MHz and (ii) the total power absorbed by the electrons increases by a factor of about 3.6 at the same frequency, whereas the ion velocity at the electrodes increases by about 50 \%. According to equation (\ref{Model}) the increase of the ion velocity alone would lead to a lower density, while the decrease of $\varepsilon_{\rm{e}}$ and the increase of $S_{\rm{abs}}$ cause a density increase. The latter two effects are dominant and finally cause the sudden increases of $n_{\rm{i,el}}$ and $\bar{n}_{\rm{e}}$. $\varepsilon_{\rm{c}}$ remains almost constant ($\approx 27$ eV) as a function of the frequency. Locally a small increase ($\approx 2$ eV) of $\varepsilon_{\rm{c}}$ is observed close to the density jump. This increase is a result of a drop of the mean electron energy around 60 MHz. 
This can be directly seen from the Electron Energy Probability Function (EEPF), which will be discussed later (cf. figure \ref{EEDF}). For frequencies well above 60 MHz the mean electron energy increases again and, thus, $\varepsilon_{\rm{c}}$ decreases slightly.

\begin{figure}[h!]
\begin{center}
\begin{tabular}{cc}
  \includegraphics[width=0.5\textwidth]{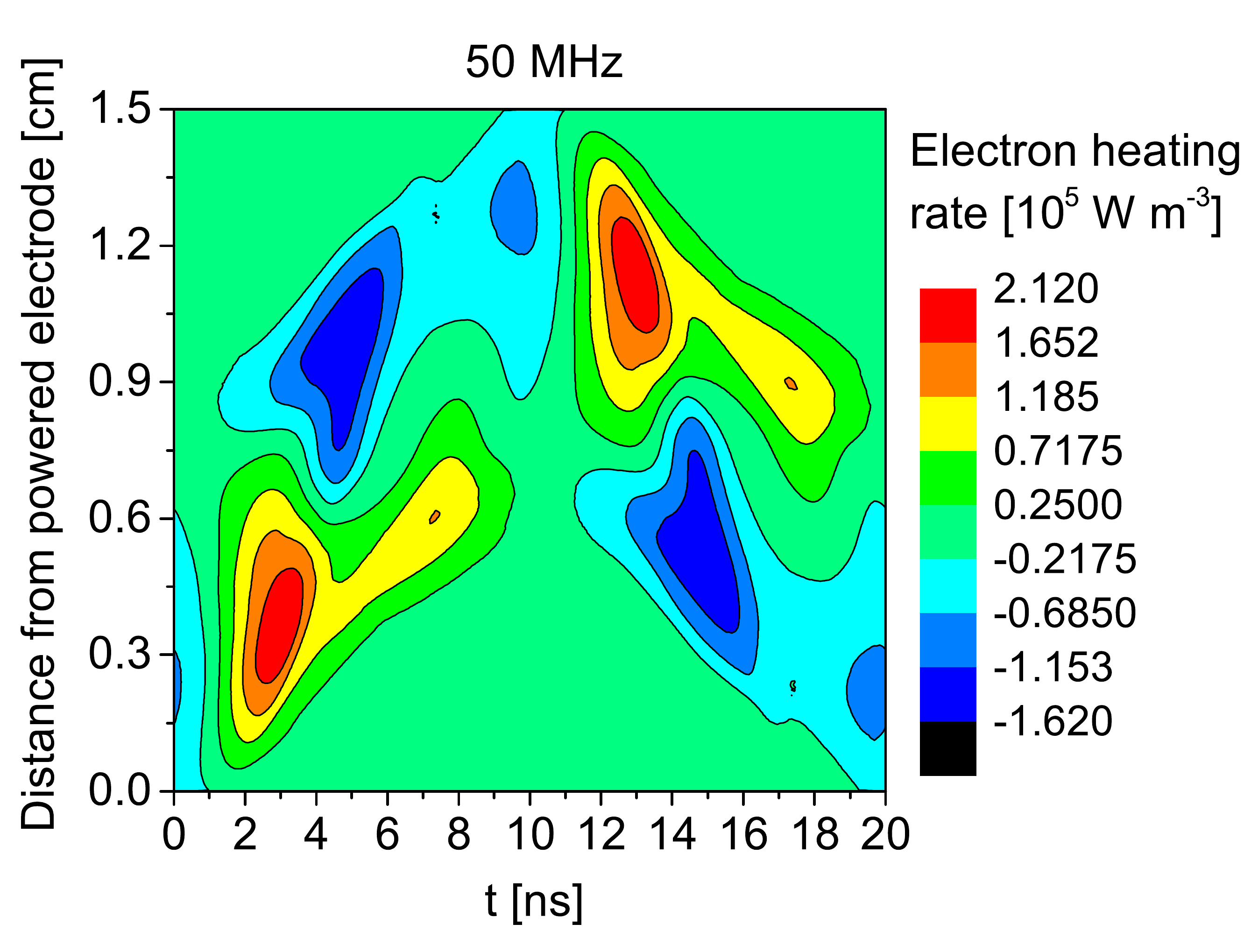}
 &
  \includegraphics[width=0.5\textwidth]{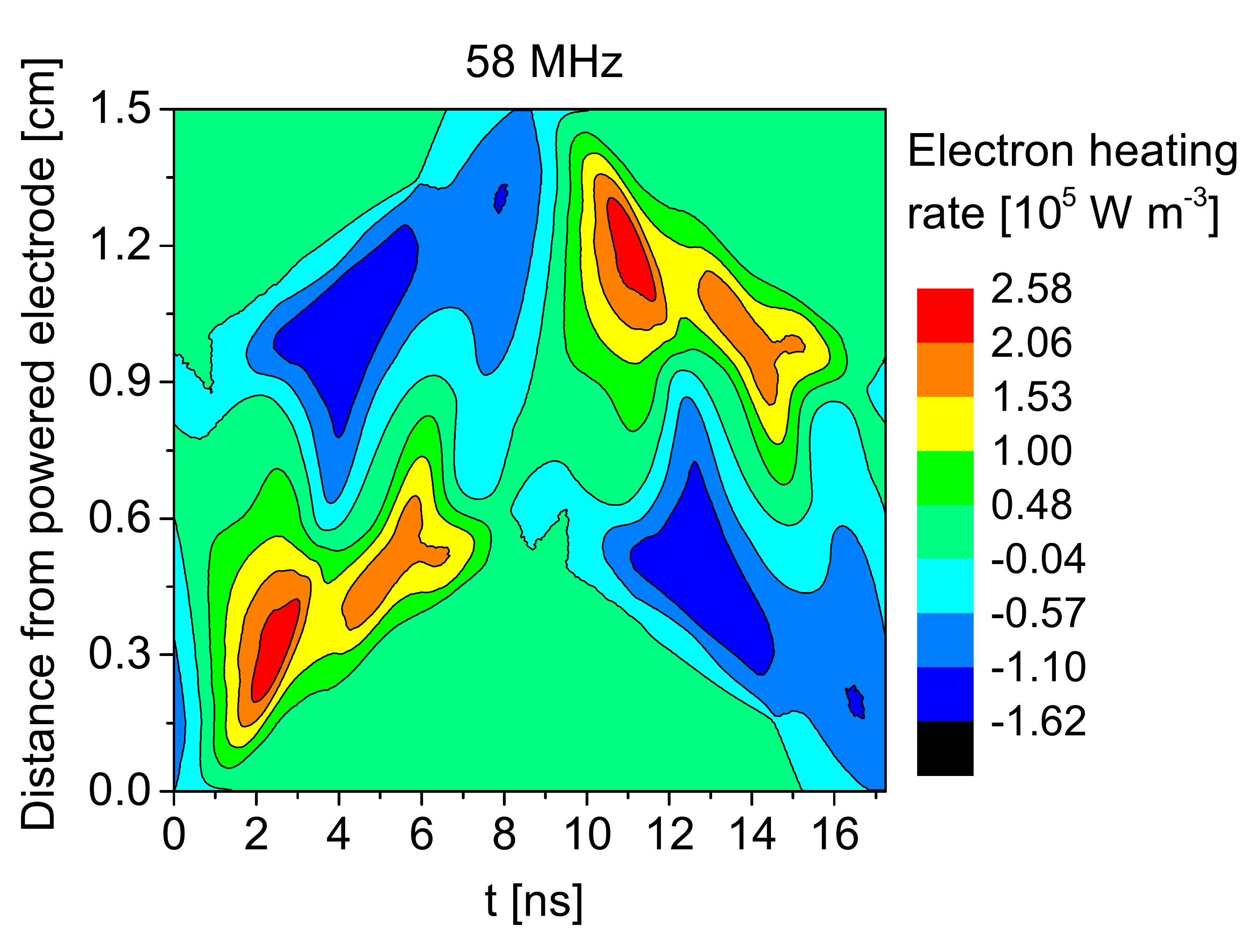}
 \\
  \includegraphics[width=0.5\textwidth]{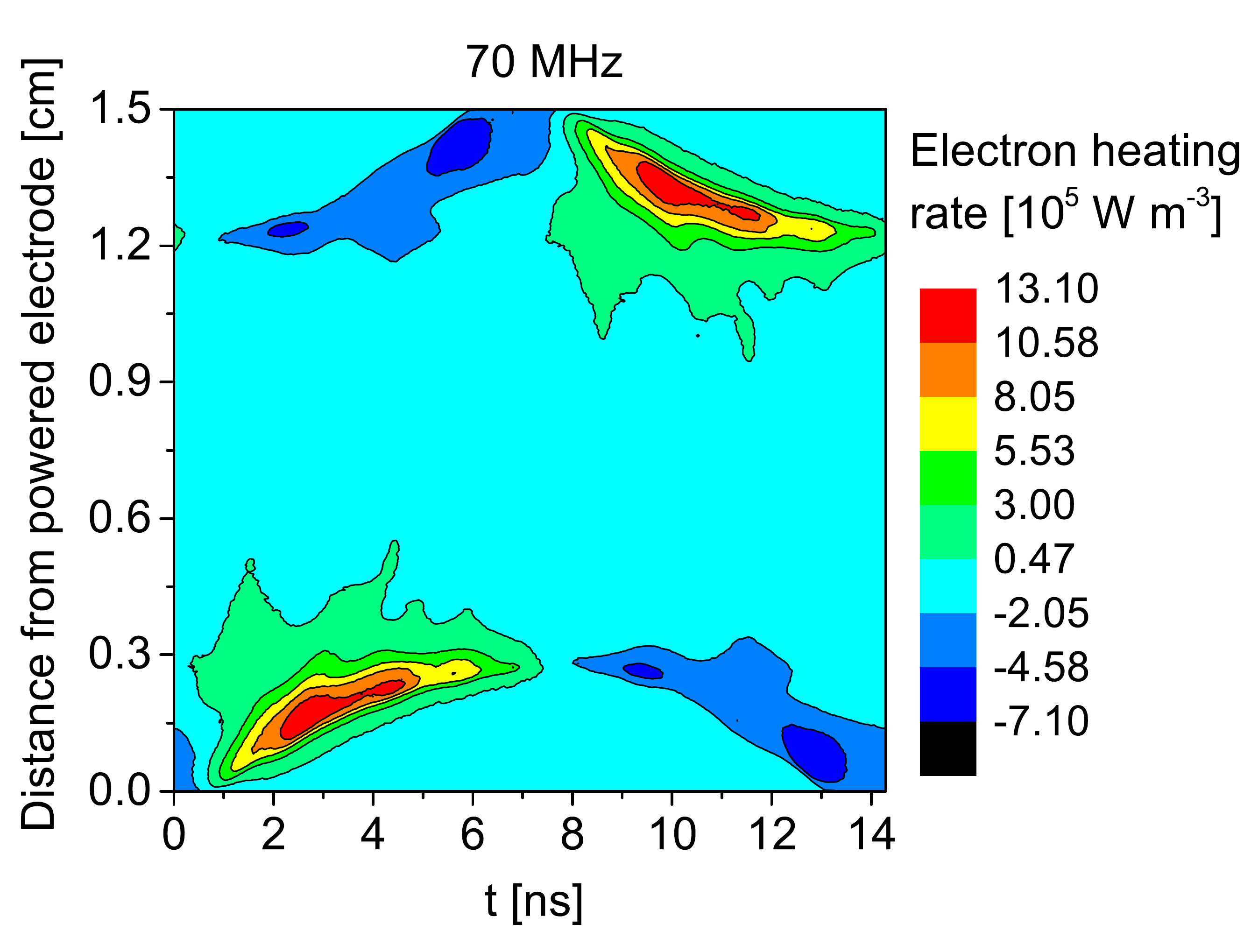}
 &
  \includegraphics[width=0.5\textwidth]{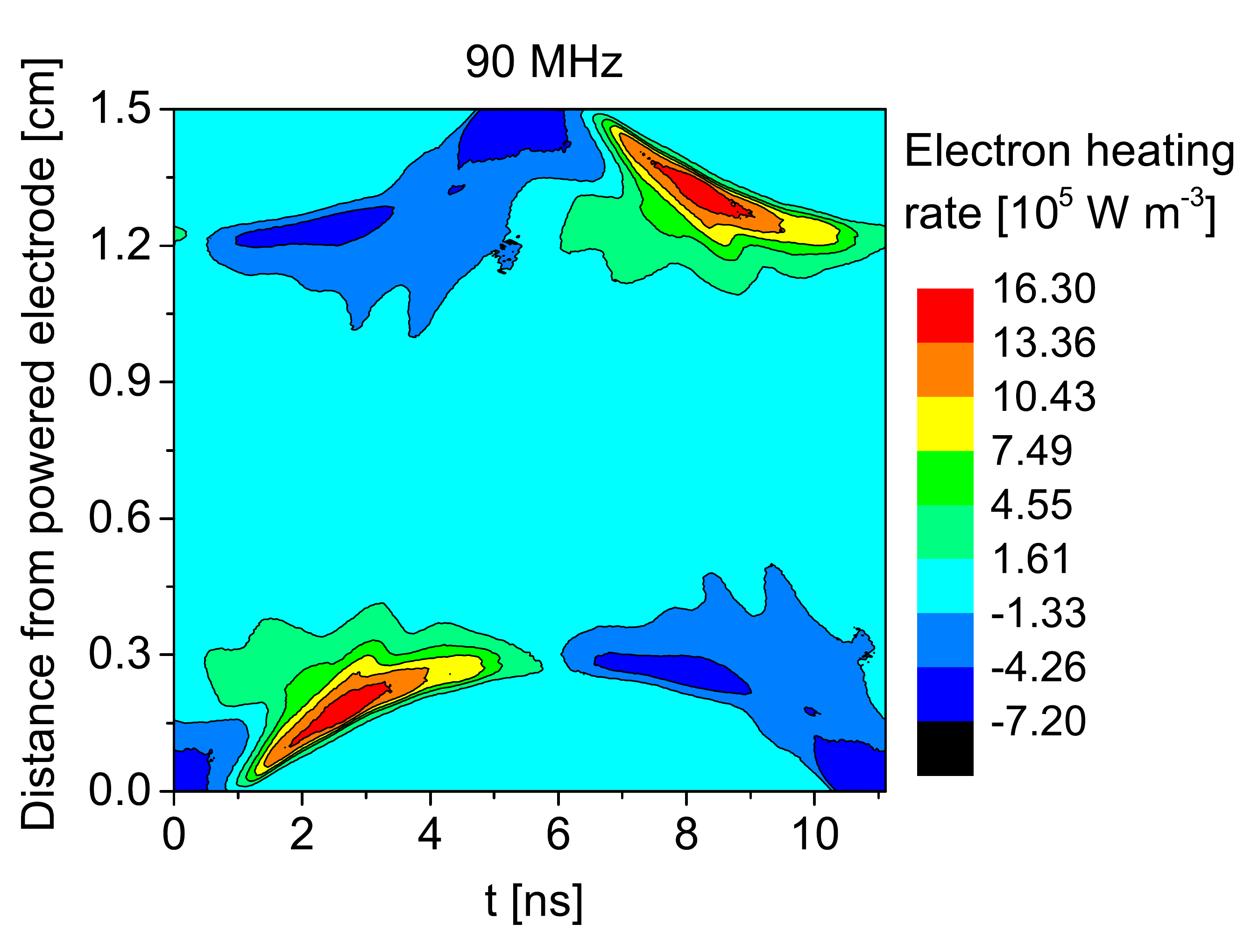}
 \\
\end{tabular}
\caption{Spatio-temporal plots of the electron heating rate within one RF period between 50 MHz and 90 MHz obtained from the simulation. Note that the density jump occurs at 60 MHz and that the duration of one RF period is different in each plot due to the different driving frequencies. Conditions: $p = 1.3$ Pa, $d = 1.5$ cm, $\phi_{0} = 150$ V, $s = 1$, and $\gamma = 0$.}
\label{Heating}
\end{center}
\end{figure}

\begin{figure}[h!]
\begin{center}
\begin{tabular}{cc}
  \includegraphics[width=0.5\textwidth]{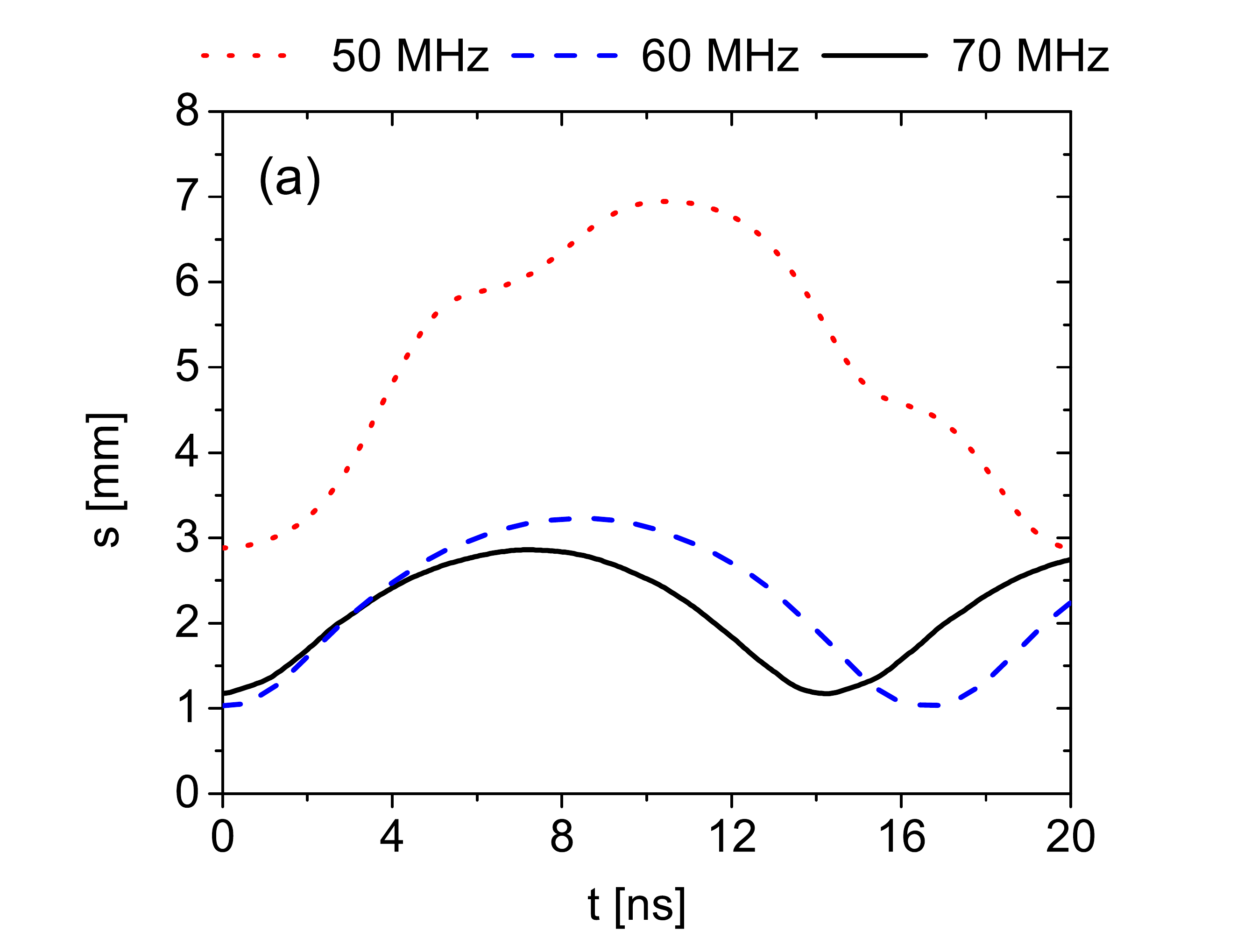}
 &
  \includegraphics[width=0.5\textwidth]{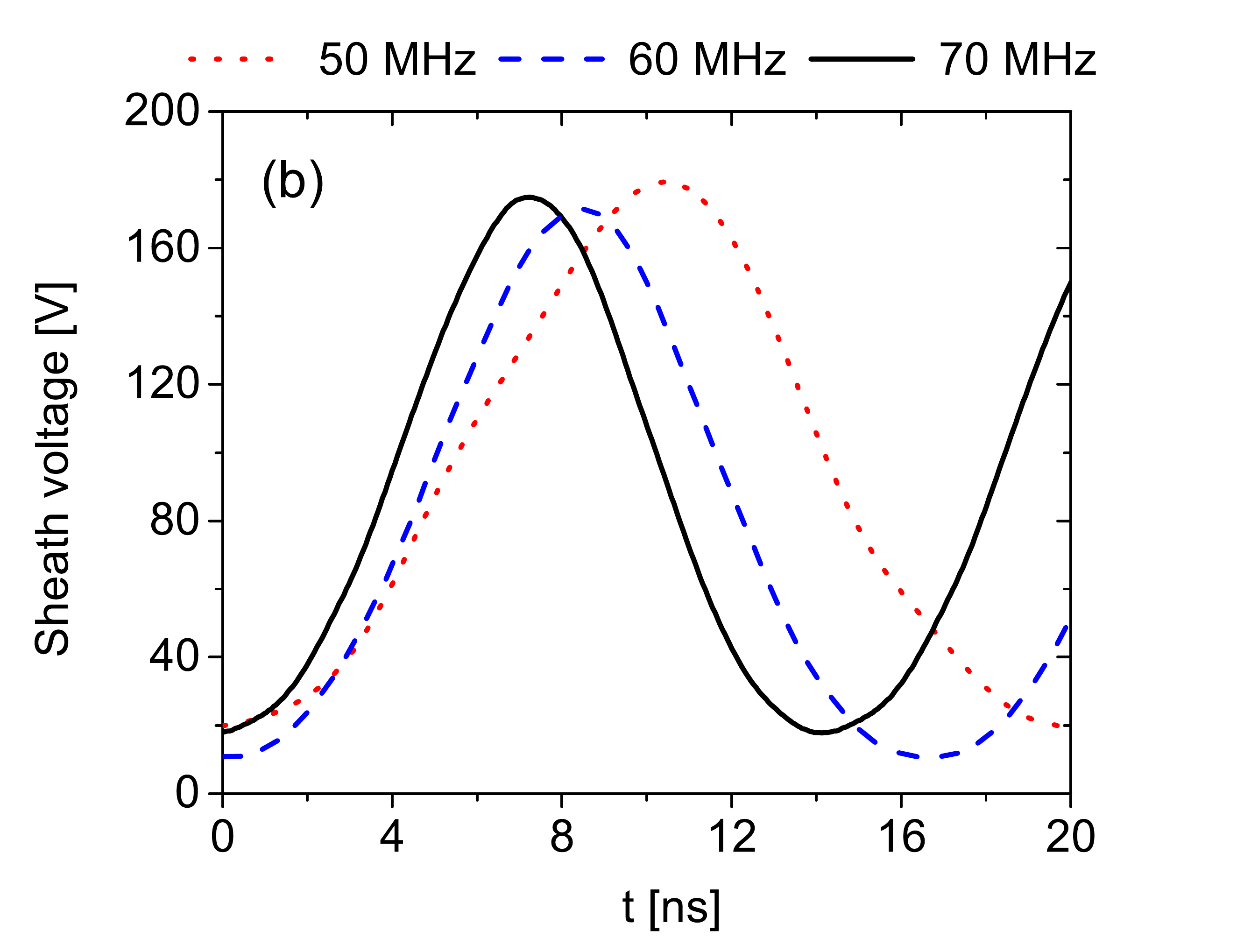}
 \\
\end{tabular}
\caption{Sheath width (left) and sheath voltage (right) at the powered electrode as a function of time within one RF period. Note that the density jump occurs at 60 MHz and that the duration of one RF period is different for each driving frequency. Conditions: $p = 1.3$ Pa, $d = 1.5$ cm, $\phi_{0} = 150$ V, $s = 1$, and $\gamma = 0$.}
\label{Sheath}
\end{center}
\end{figure}

The step-like behavior of $\varepsilon_{\rm{e}}$ and $S_{\rm{abs}}$ is related to a sudden electron heating mode transition around 60 MHz such as shown in figure \ref{Heating}. At high driving frequencies and for a given driving voltage amplitude the ion density is high and the sheaths are small. Thus, the sheaths expand slowly as shown in the left plot of figure \ref{Sheath} for 70 MHz as an example. Here, the sheath width is calculated as a function of time within the RF period according to a criterion proposed by Brinkmann\cite{SCrit}. The discharge is operated in the classical $\alpha$-mode and one electron beam is generated at each electrode within each RF period. Most of these beam electrons hit the opposing sheath, when the local sheath voltage is high and are reflected back into the bulk, i.e. the confinement is good and $\varepsilon_{\rm{e}}$ is low.  Decreasing the frequency from 70 MHz to 60 MHz causes the ion density at the electrodes to decrease moderately (see figure \ref{Density}). This leads to a bigger sheath and a faster sheath expansion at 60 MHz compared to 70 MHz (see left plot of figure \ref{Sheath}). In the frequency range below 60 MHz the sheaths expand so fast that the electrons located at the sheath edge cannot react and oscillations in the electron heating rate at frequencies comparable to the local electron plasma frequency are excited, similar to previous findings of Vender et al. \cite{VenderFieldRev}. Based on Ku et al. \cite{Annaratone,Ku,Ku2} these oscillations might be considered to be the consequences of a local excitation of the Plasma Parallel Resonance. This effect is noticeable for the low density resonant heating mode. In this mode two electron beams are generated at each electrode during one phase of sheath expansion. Accordingly, at low driving frequencies two separated maxima in the electron heating rate are observed at each electrode during sheath expansion resulting in the generation of two subsequent electron beams during one phase of sheath expansion. While the first beam is well confined, the second beam hits the opposing sheath during its collapse. Many energetic electrons are therefore lost at the electrode, i.e., the confinement of energetic electrons is poor and $\varepsilon_{\rm{e}}$ is high. This effect is self-amplifying and keeps the discharge in this low density resonant heating mode. At high driving frequencies one maximum in the electron heating rate is observed at each electrode during sheath expansion resulting in the generation of a single electron beam. 


\begin{figure}[h!]
\begin{center}
\begin{tabular}{cc}
  \includegraphics[width=0.5\textwidth]{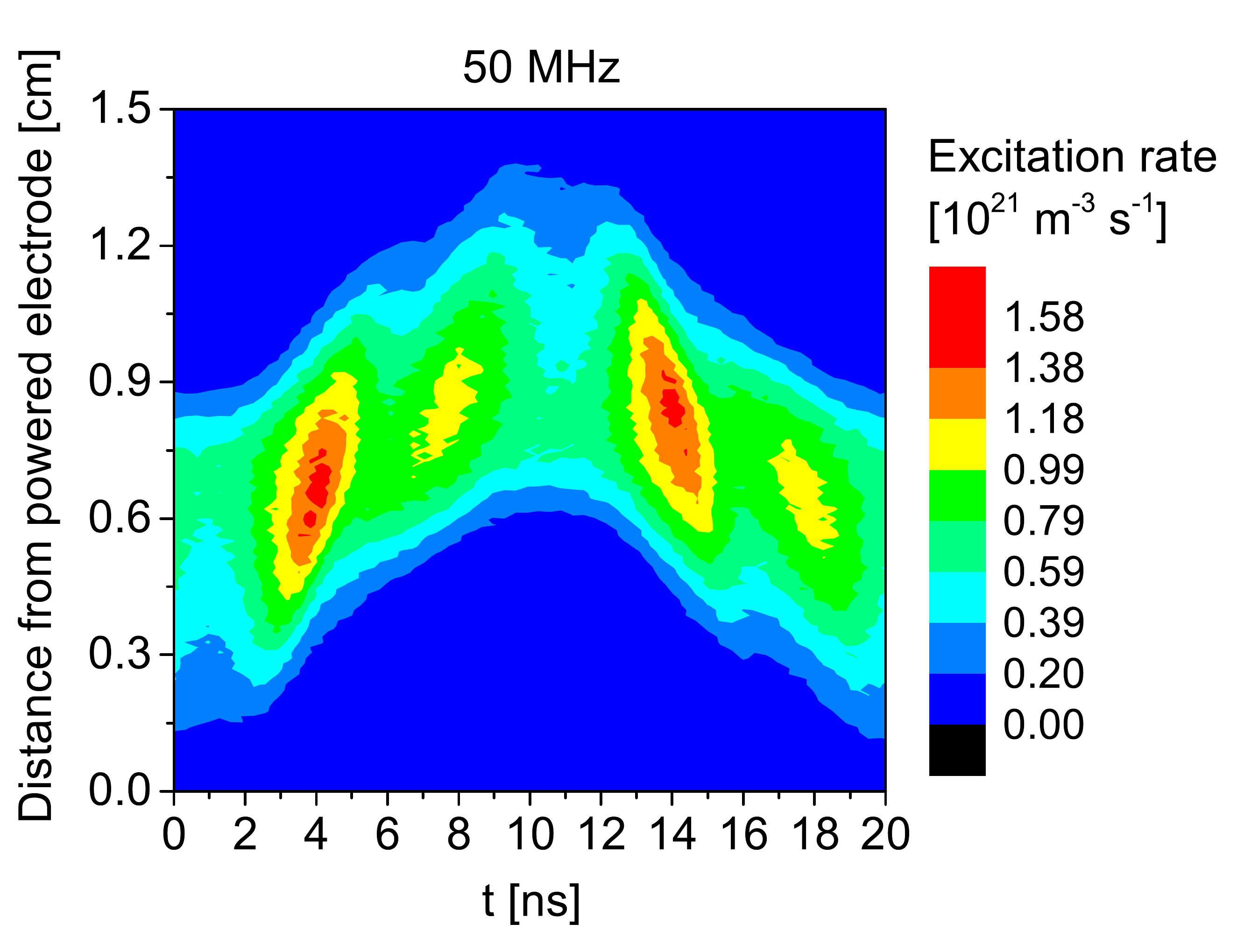}
 &
  \includegraphics[width=0.5\textwidth]{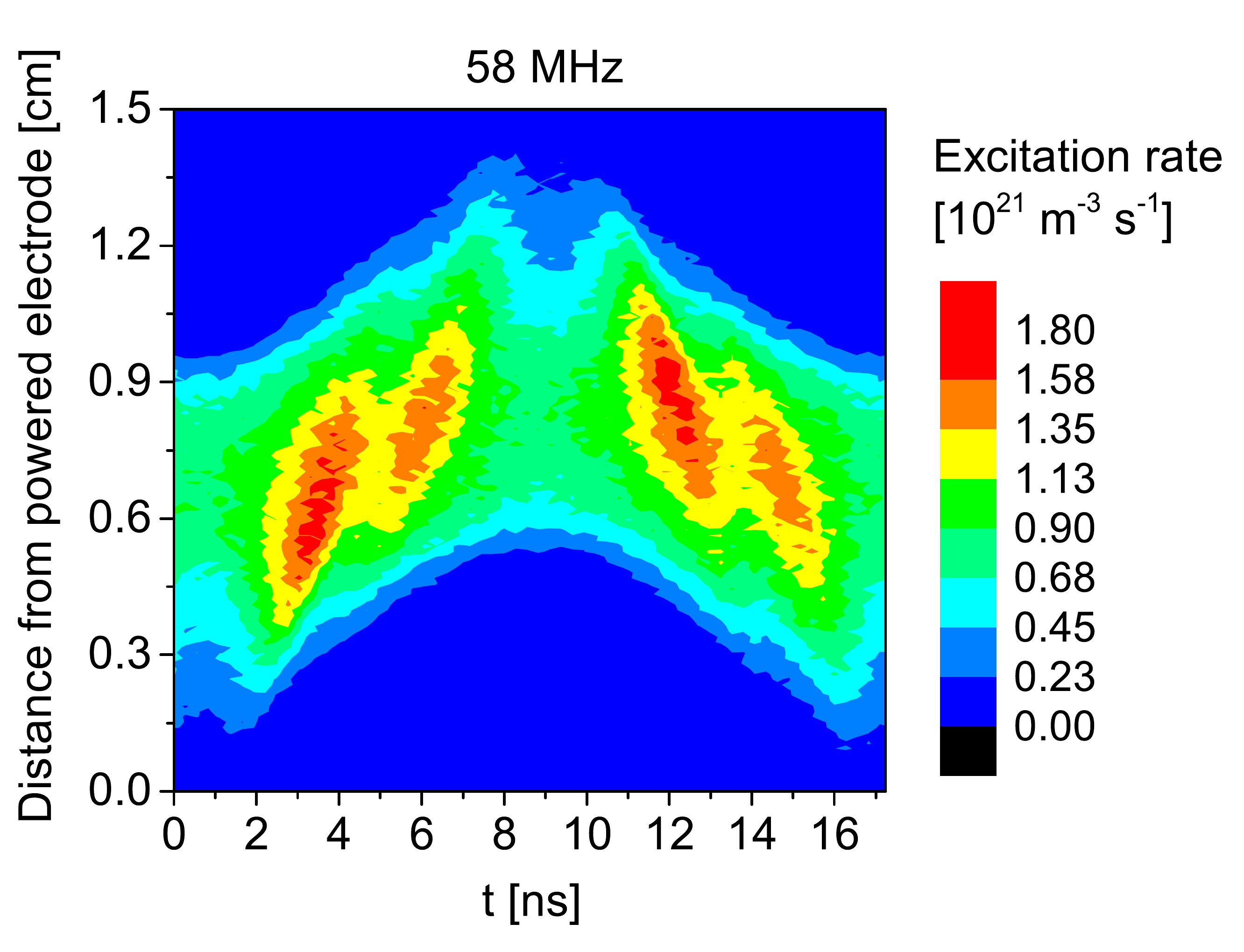}
 \\
  \includegraphics[width=0.5\textwidth]{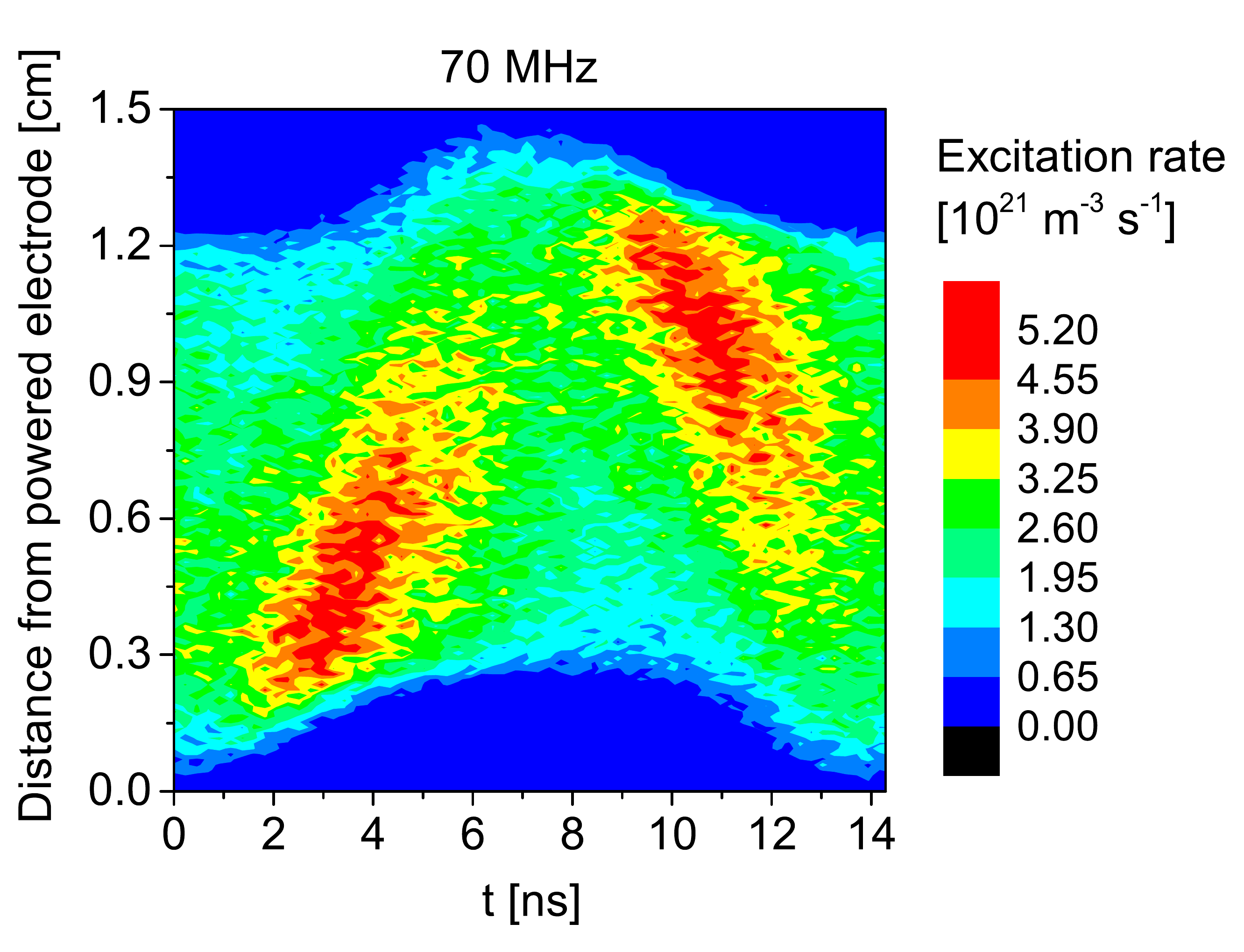}
 &
 \includegraphics[width=0.5\textwidth]{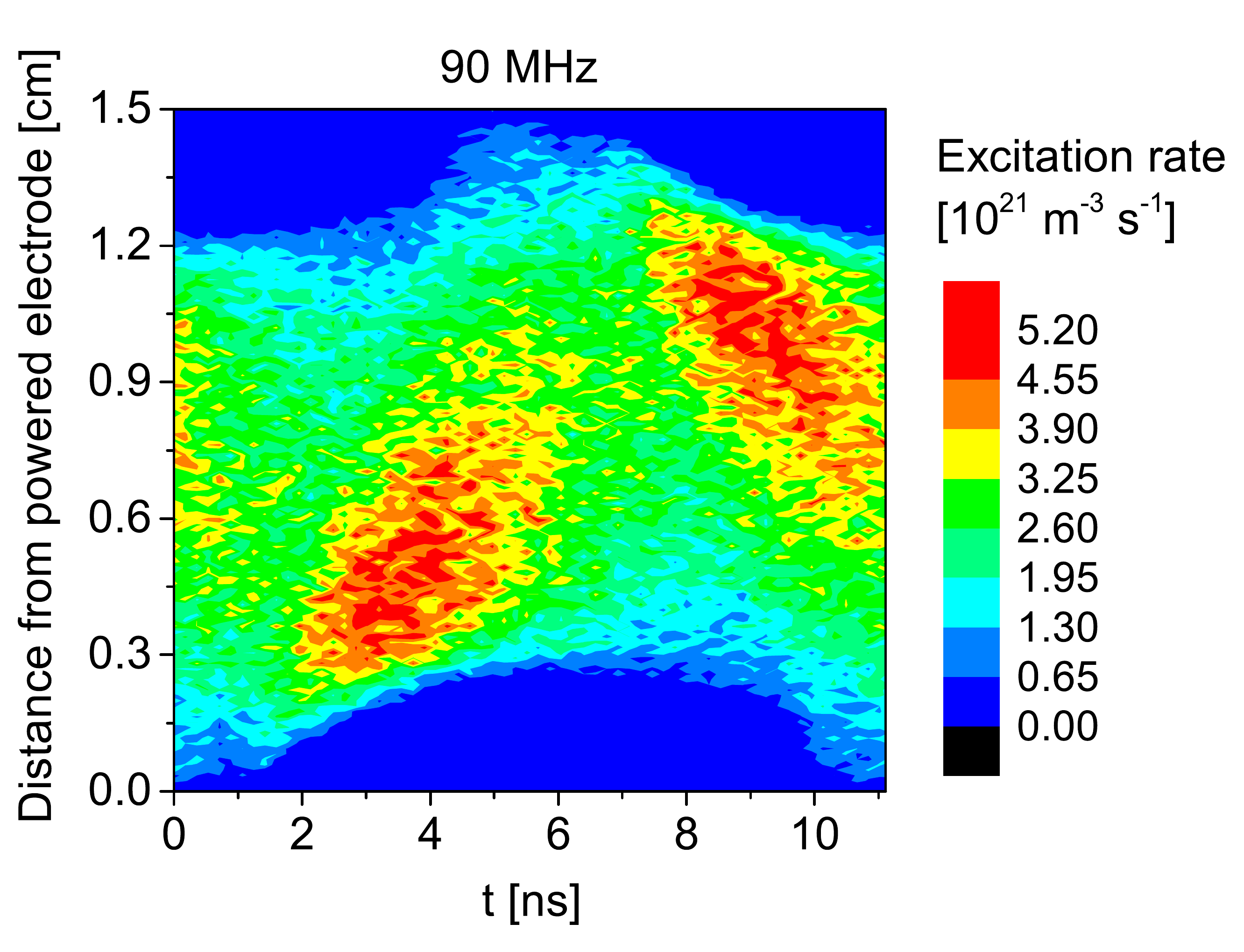}
 \\
\end{tabular}
\caption{Spatio-temporal plots of the electron impact excitation rate within one RF period for different driving frequencies. Note that the density jump occurs at 60 MHz and that the duration of one RF period is different in each plot due to the different driving frequencies. Conditions: $p = 1.3$ Pa, $d = 1.5$ cm, $\phi_{0} = 150$ V, $s = 1$, and $\gamma = 0$.}
\label{Excitation}
\end{center}
\end{figure}

The different number of electron beams generated per sheath expansion phase in the high and low density modes are clearly visible in the spatio-temporal plots of the electron impact excitation rates shown in figure \ref{Excitation}. The excitation rate can be measured experimentally by Phase Resolved Optical Emission Spectroscopy (PROES) \cite{PROES,UCZFieldRev,GrahamFieldRev}, which might be used as a diagnostic to verify these simulation results in the future. 

\begin{figure}[h!]
\begin{center}
\begin{tabular}{cc}
  \includegraphics[width=0.5\textwidth]{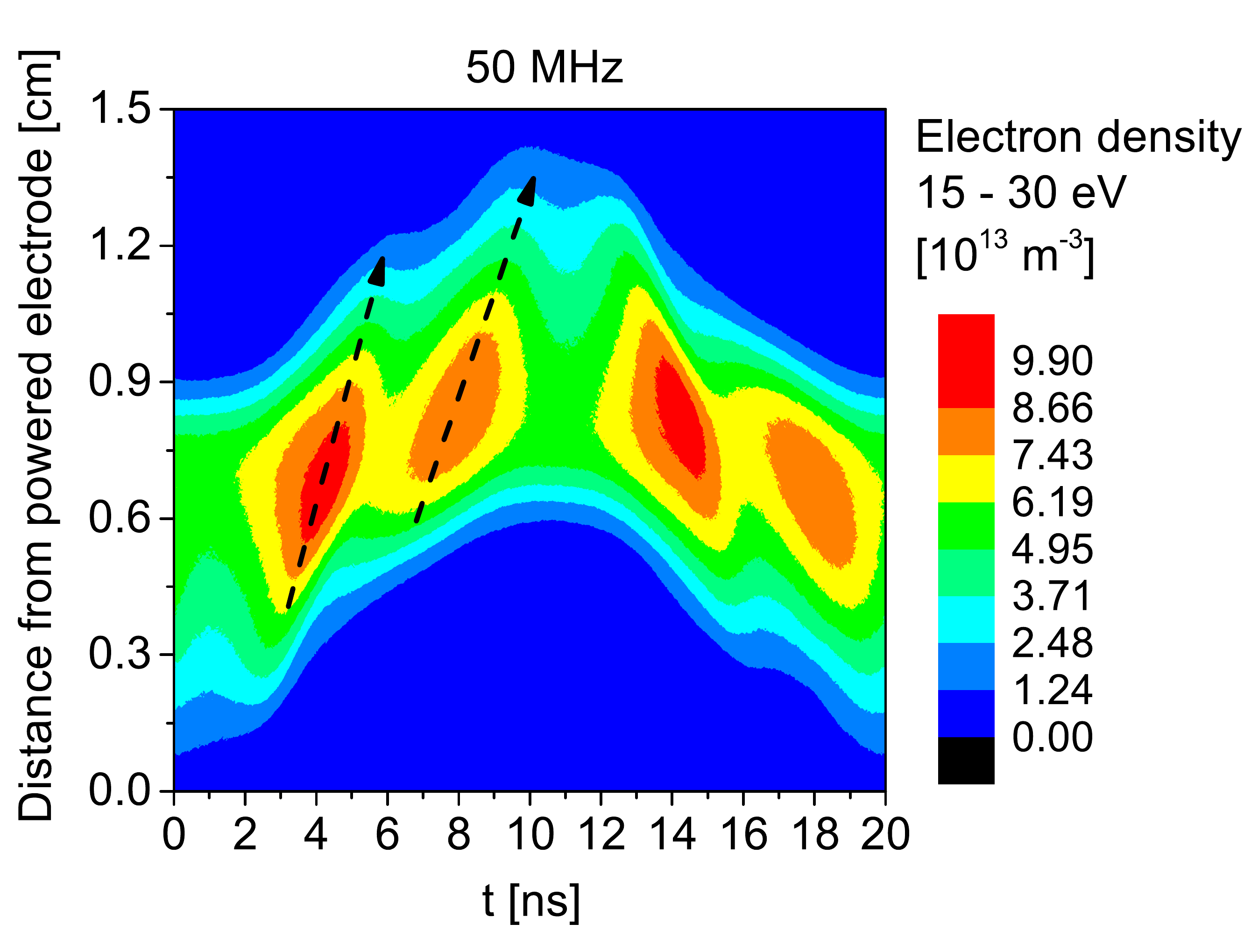}
 &
  \includegraphics[width=0.5\textwidth]{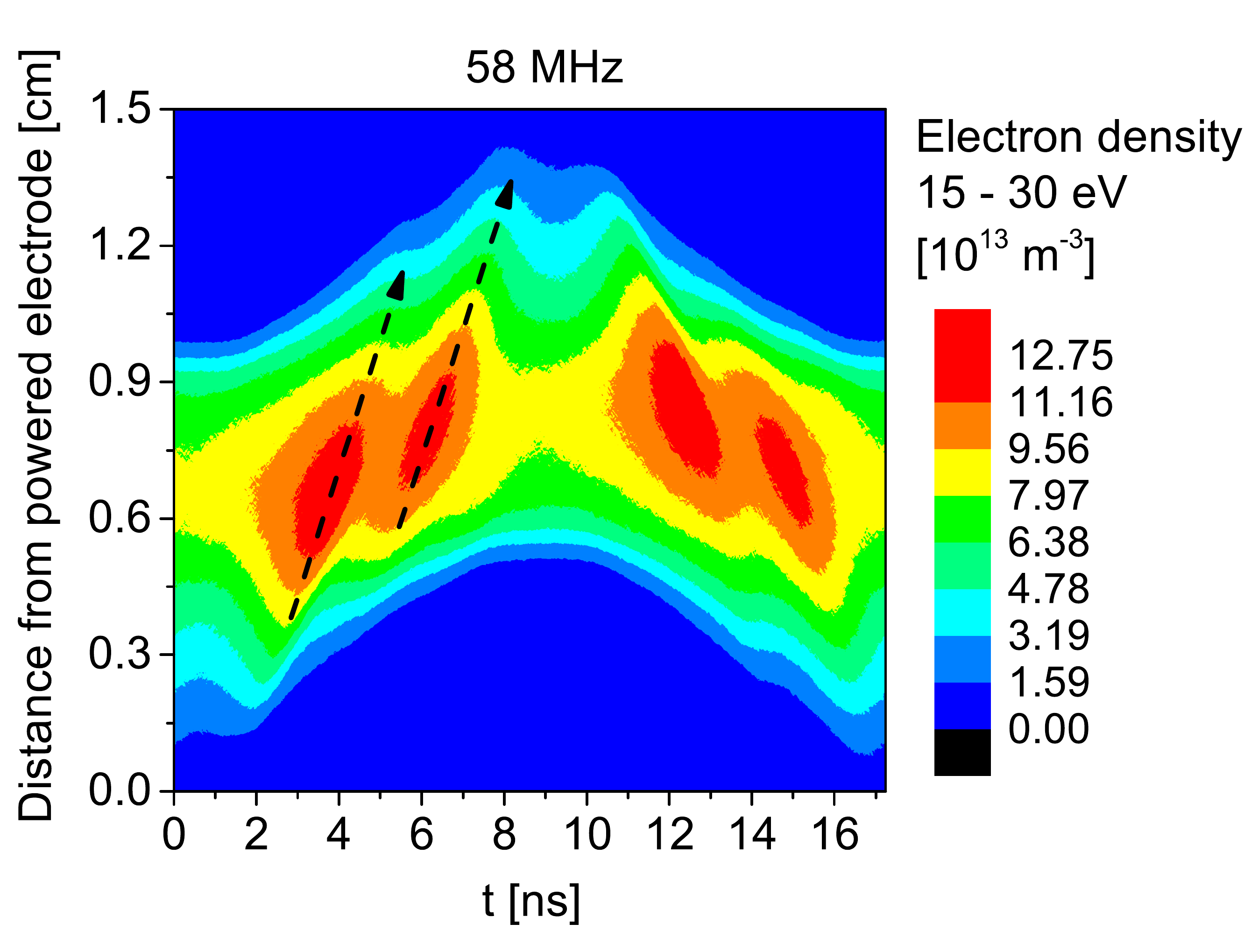}
 \\
  \includegraphics[width=0.5\textwidth]{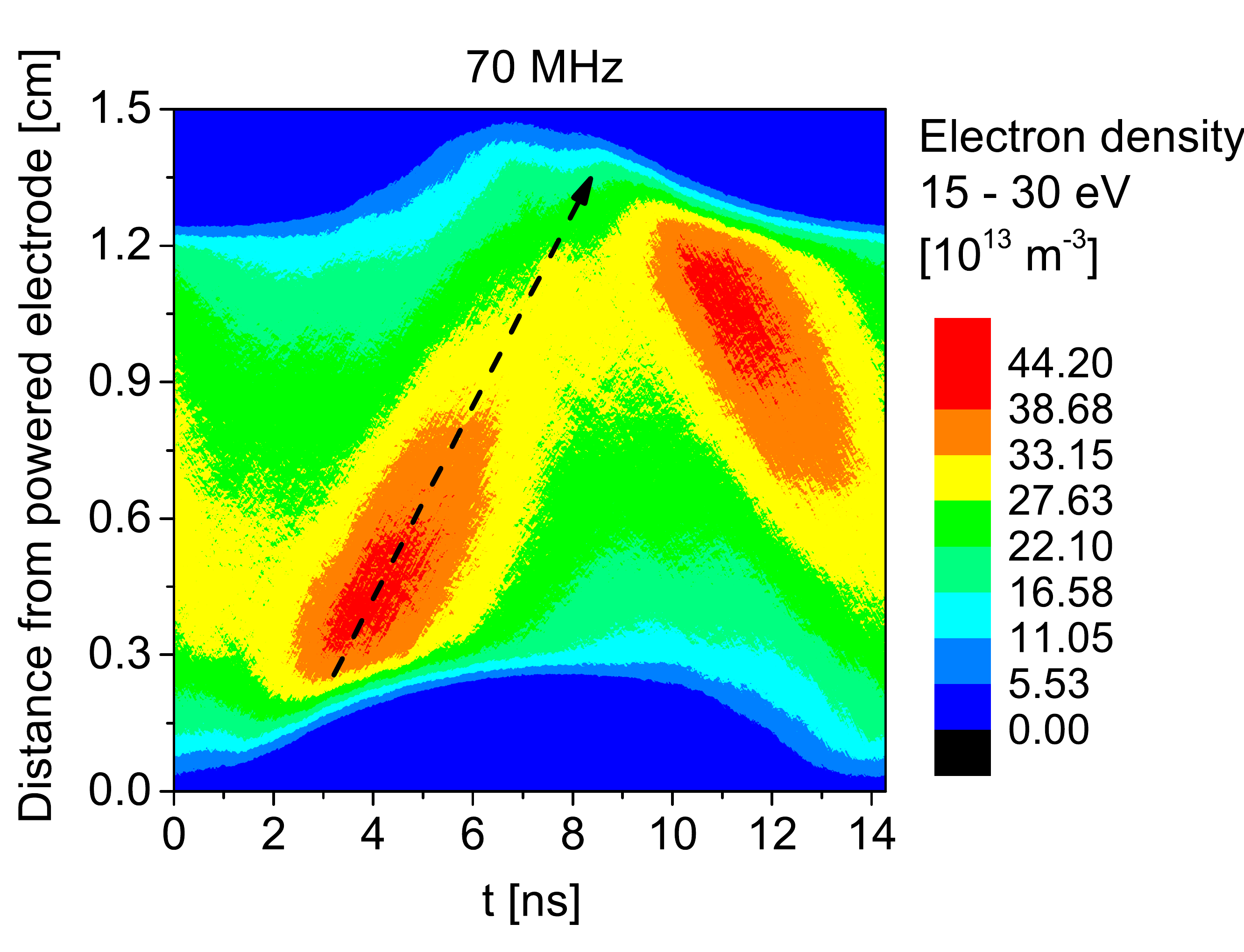}
 &
  \includegraphics[width=0.5\textwidth]{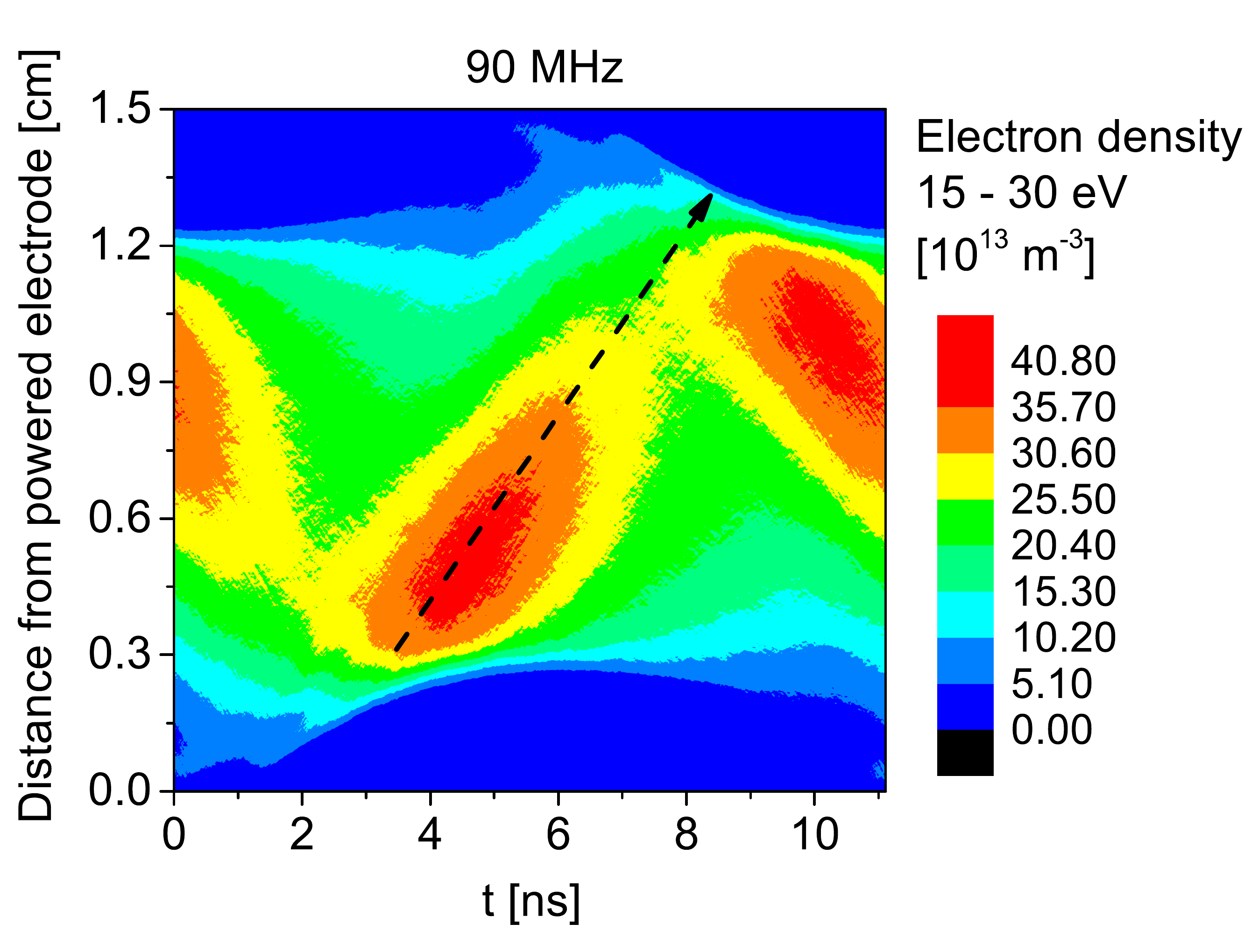}
 \\
\end{tabular}
\caption{Spatio-temporal plots of the density of electrons with energies between 15 and 30 eV. Conditions: $p = 1.3$ Pa, $d = 1.5$ cm, $\phi_{0} = 150$ V, $s = 1$, and $\gamma = 0$. The arrows indicate the path of beam electrons.}
\label{HighEnergy1}
\end{center}
\end{figure}

\begin{figure}[h!]
\begin{center}
\begin{tabular}{cc}
  \includegraphics[width=0.5\textwidth]{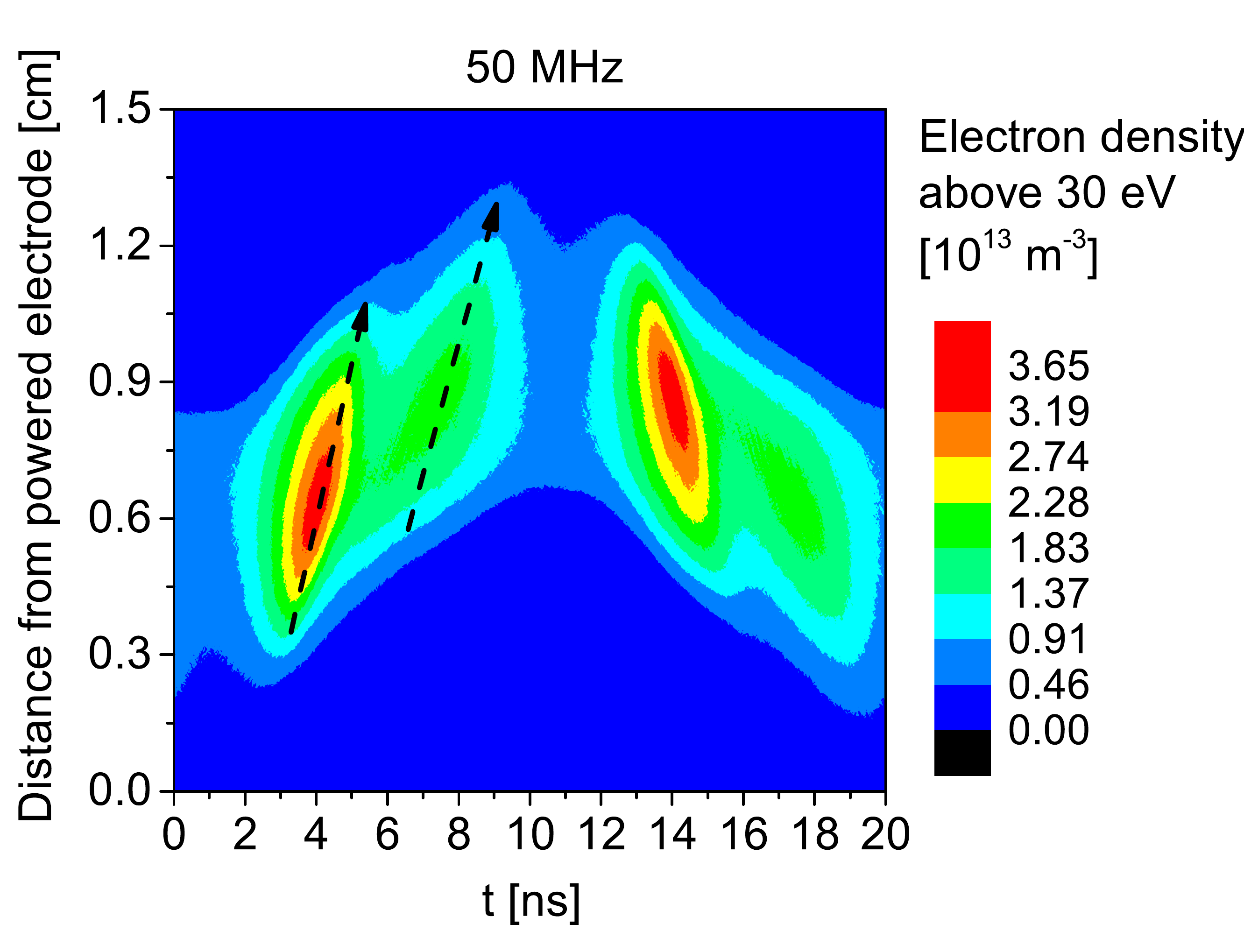}
 &
  \includegraphics[width=0.5\textwidth]{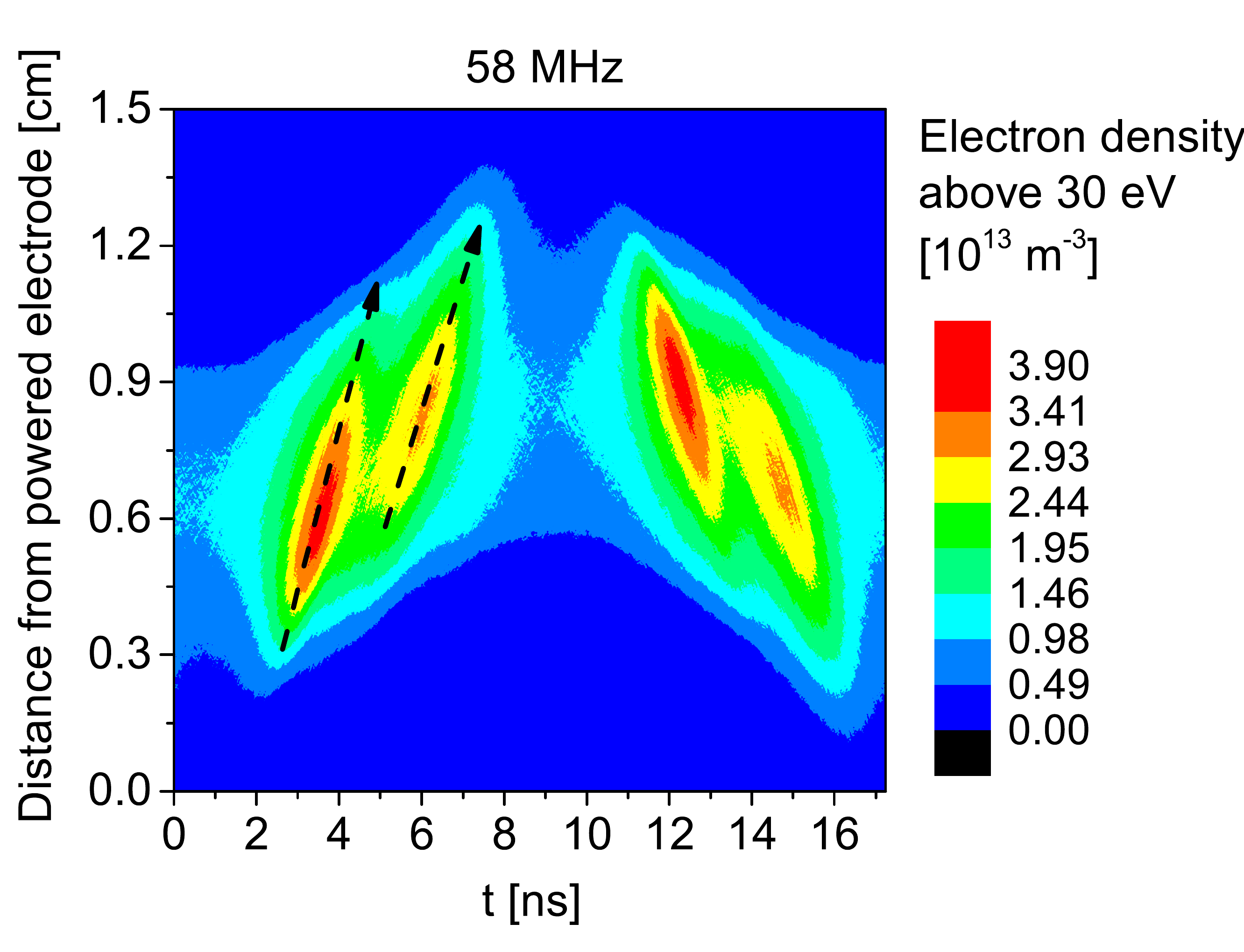}
 \\
  \includegraphics[width=0.5\textwidth]{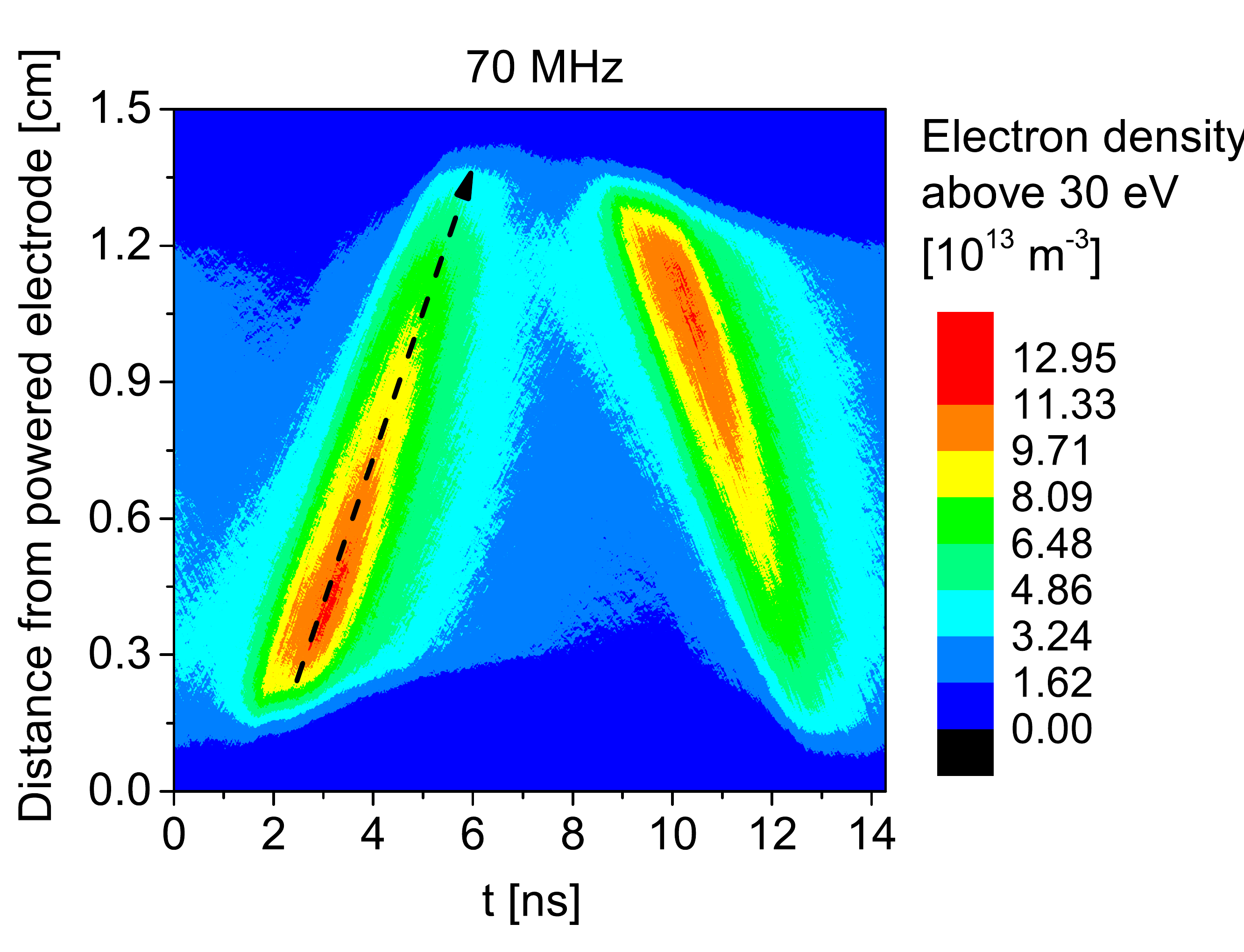}
 &
  \includegraphics[width=0.5\textwidth]{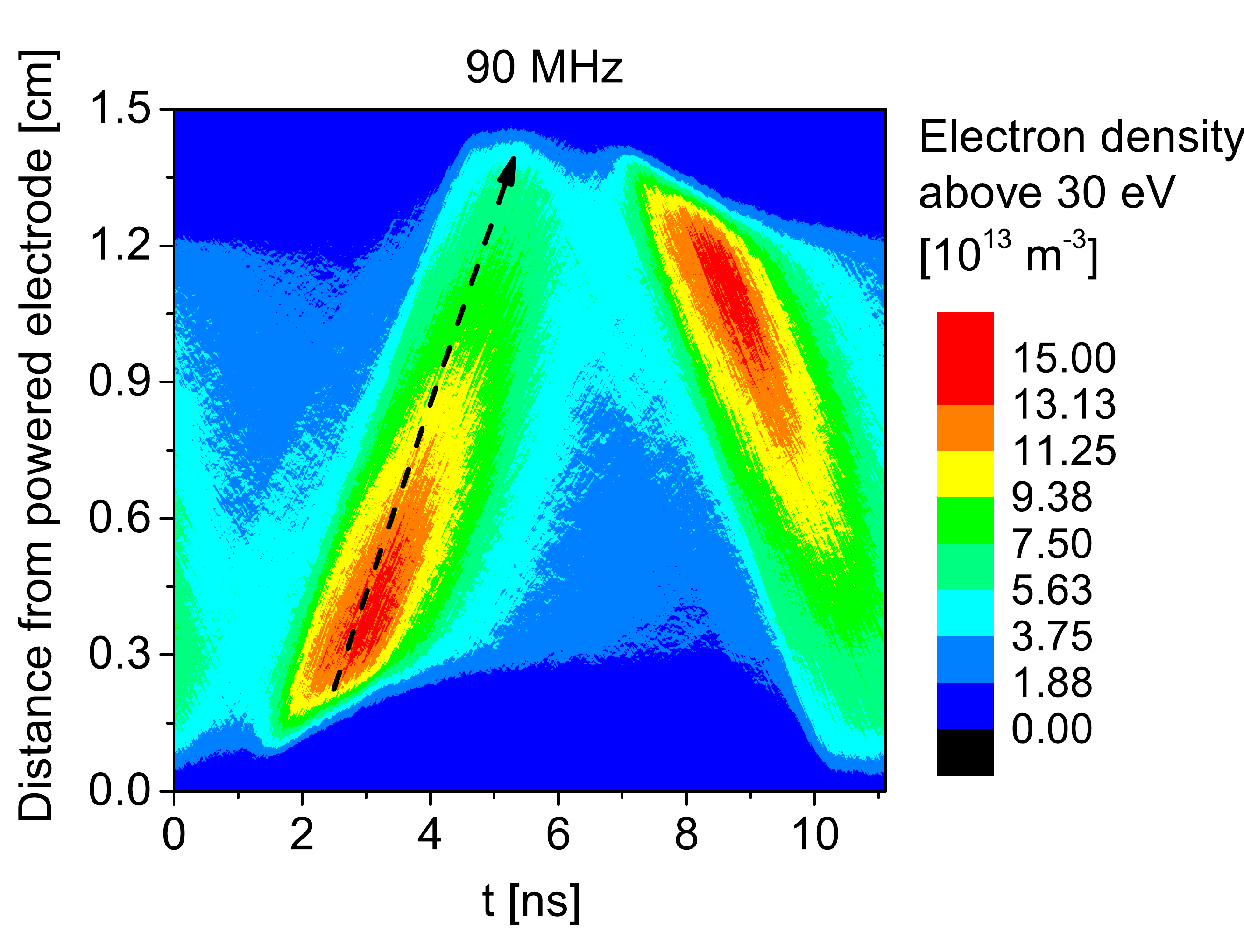}
 \\
\end{tabular}
\caption{Spatio-temporal plots of the density of electrons with energies above 30 eV. Conditions: $p = 1.3$ Pa, $d = 1.5$ cm, $\phi_{0} = 150$ V, $s = 1$, and $\gamma = 0$. The arrows indicate the path of beam electrons.}
\label{HighEnergy2}
\end{center}
\end{figure}

At a pressure of 1.3 Pa the ratio of the electron mean free path, $\lambda_{\rm{m}}$, and the electrode gap, $d$, is $\lambda_{\rm{m}}/d \approx 2.3$, i.e., most beam electrons reach the opposing sheath edge without any collision. However, the electron beams generated during the phases of sheath expansion are not mono-energetic. This is illustrated in figures \ref{HighEnergy1} and \ref{HighEnergy2} that show spatio-temporal plots of the density of electrons with energies between 15 and 30 eV as well as above 30 eV, respectively. These results are obtained from the simulation under the conditions of the base case. Low energetic beam electrons (within the finite energy range of 15 eV - 30 eV) hit the opposing sheath at different phases for different driving frequencies. This stems from the shortening of the RF period at higher driving frequencies, while the electrode gap and the electron energy range shown in figure \ref{HighEnergy1} remain constant. These low energy beam electrons hit the opposing sheath during its collapse only at low frequencies. In this case the high $\varepsilon_{e}$ keeps the discharge in the low density resonant heating mode. Generally, an electron beam is generated at one electrode during sheath expansion within a relatively narrow region in space and time, but spreads on its way to the opposing electrode resulting in a wider time interval of electron beam interaction with the sheath at the opposing electrode compared to the electrode at which it was generated.

\begin{figure}[h!]
\begin{center}
\begin{tabular}{cc}
  \includegraphics[width=0.5\textwidth]{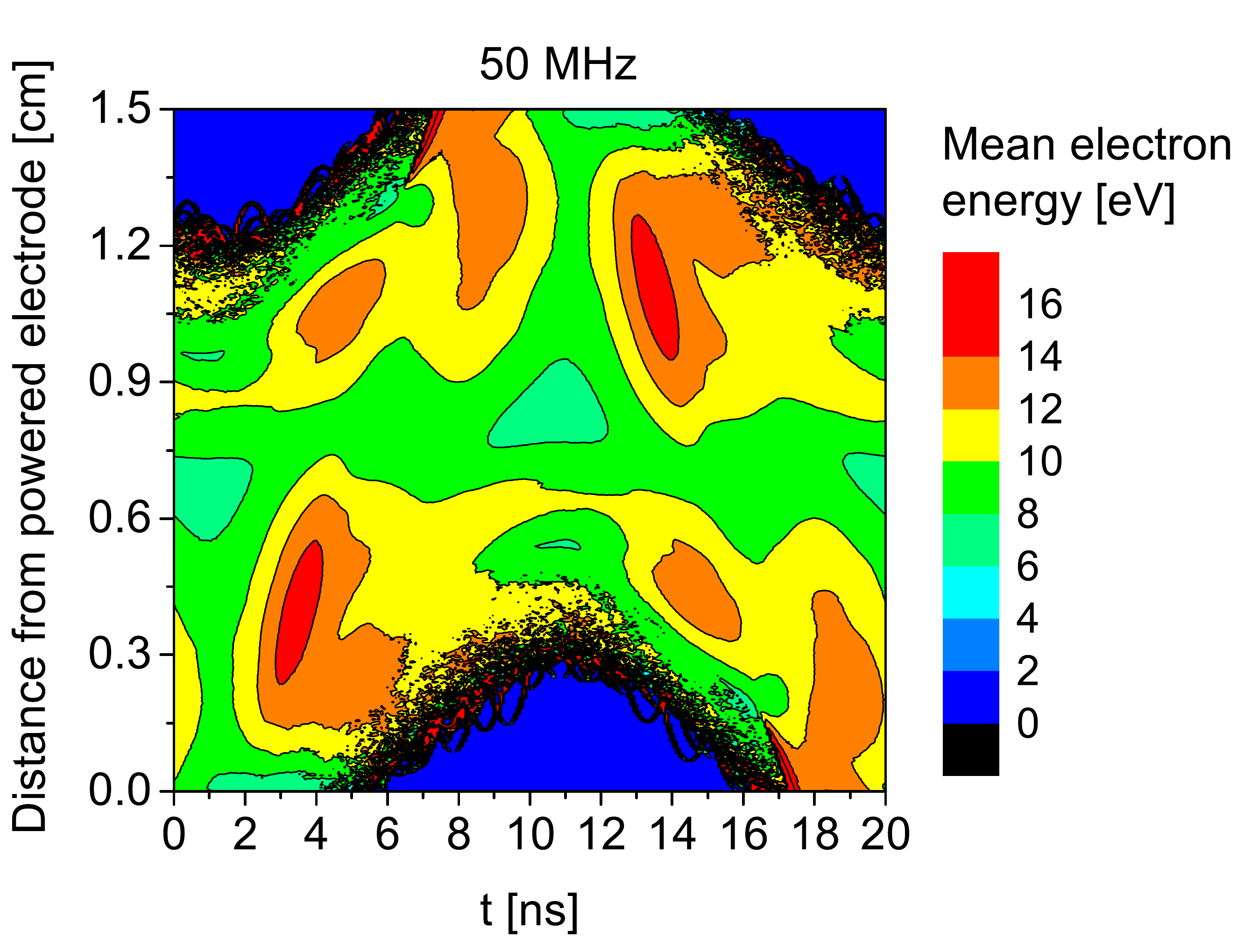}
 &
  \includegraphics[width=0.5\textwidth]{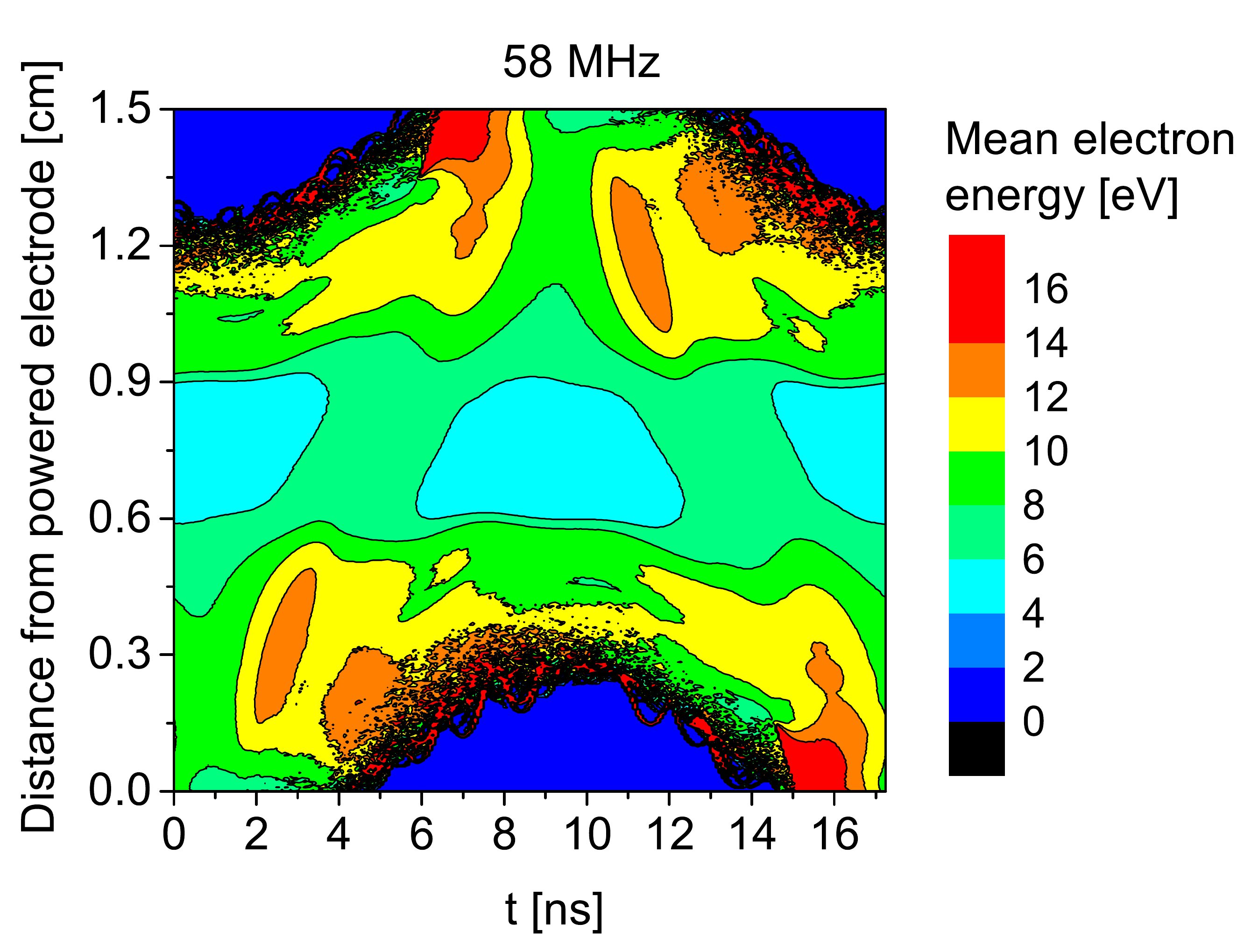}
 \\
  \includegraphics[width=0.5\textwidth]{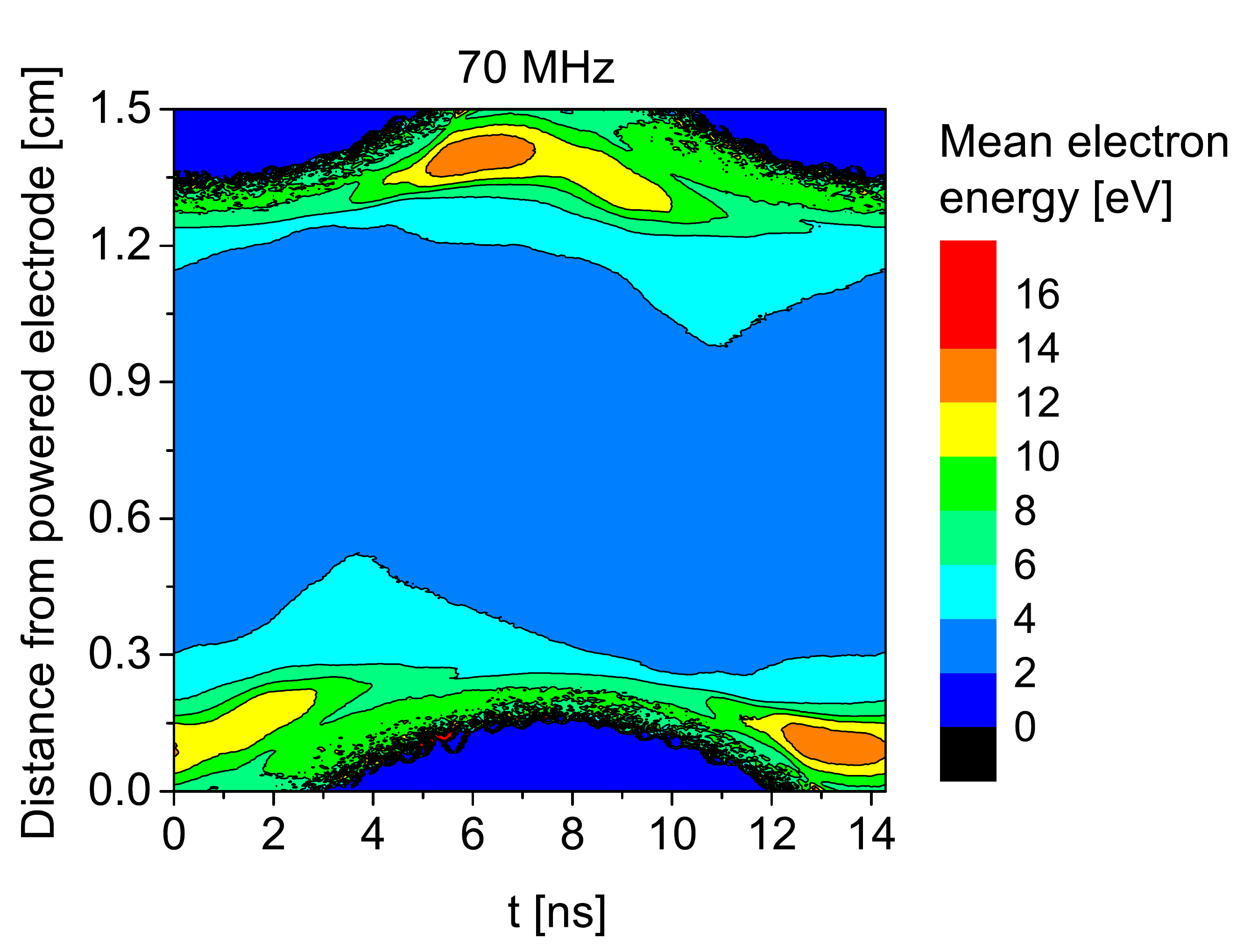}
 &
  \includegraphics[width=0.5\textwidth]{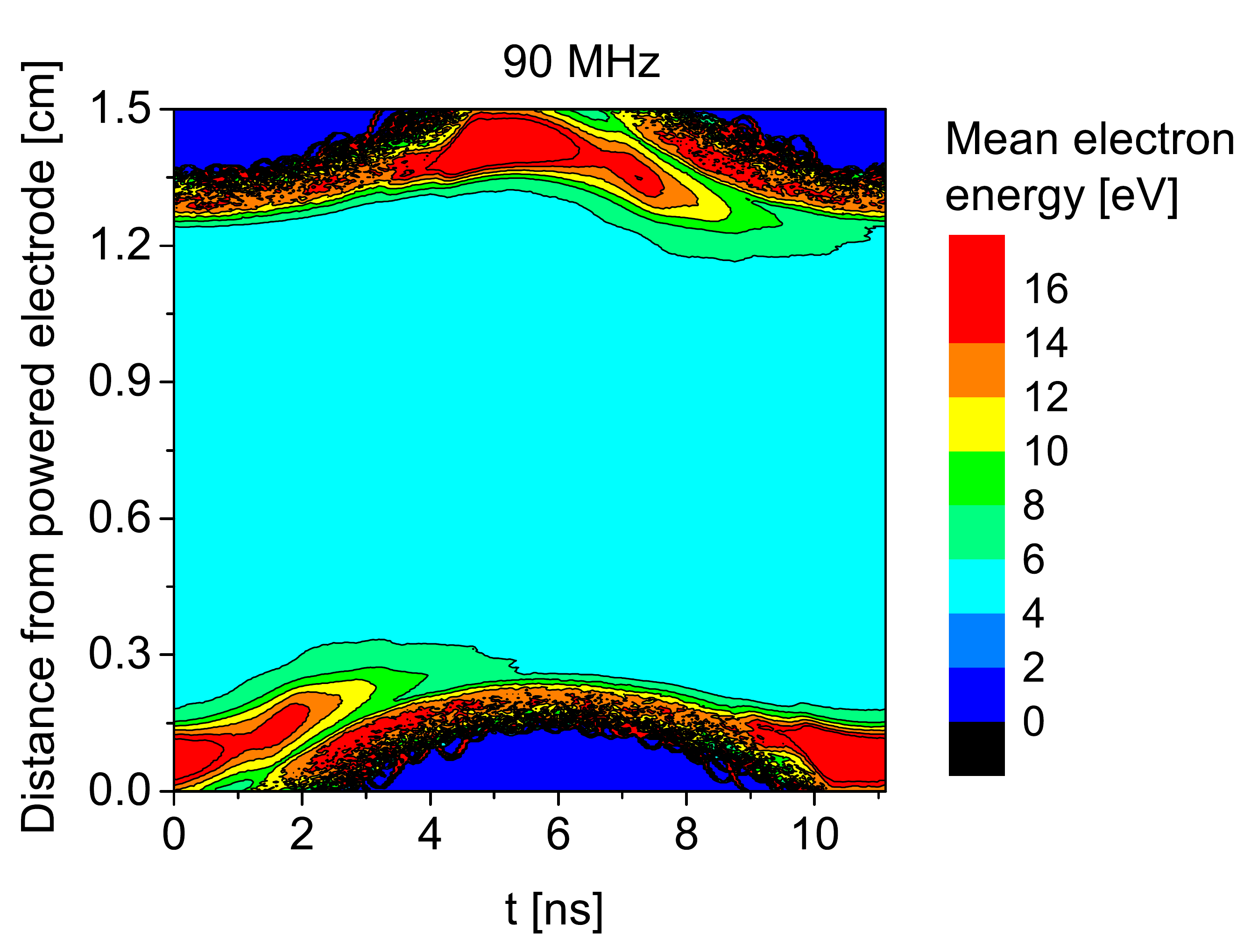}
 \\
\end{tabular}
\caption{Spatio-temporal plots of the mean electron energy within one RF period for different driving frequencies. Note that the density jump occurs at 60 MHz and that the duration of one RF period is different in each plot due to the different driving frequency. Conditions: $p = 1.3$ Pa, $d = 1.5$ cm, $\phi_{0} = 150$ V, $s = 1$, and $\gamma = 0$.}
\label{MeanEEnergy}
\end{center}
\end{figure}

The heating mode transition is initially caused by a moderate increase of the plasma density with increasing the driving frequency. Consequently, this changes the impingement dynamics of energetic beam electrons at the opposing electrode. This then leads to a reduction of the energy lost per electron lost at the electrode, $\varepsilon_{\rm{e}}$, and finally -- in a self-amplifying manner -- to an increase of the plasma density. This mechanism causes the step-like density change observed in figure \ref{Density} and a transition to the $\alpha$-mode. This process can equivalently be interpreted as a modulation of the confinement quality of energetic beam electrons according to equation (\ref{Model}). This crucial effect is further illustrated in figure \ref{MeanEEnergy}, which shows the mean electron energy as a function of position between the electrodes and time within one RF period in the low density regime (50 MHz and 58 MHz) and the high density regime (70 MHz and 90 MHz). Generally, the mean electron energy is strongly time and space modulated. It is maximum adjacent to the electrodes, whenever an energetic electron beam is generated by sheath expansion heating or when an already established electron beam hits the local sheath. In the plasma bulk the time modulation of the mean electron energy due to the presence of an energetic electron beam is much weaker due to the high number of cold thermal electrons in the bulk. As the ion density and thus the number of cold electrons decays strongly towards the electrodes, beam electrons affect the mean electron energy much more strongly in the edge regions. 

Under the conditions investigated here highly energetic beam electrons accelerated by sheath expansion adjacent to one electrode hit the opposing electrode around the phase of sheath collapse at all frequencies studied. This is indicated by a strong local maximum of the mean electron energy during sheath collapse in figure \ref{MeanEEnergy}. As there is no heating during sheath collapse (see figure \ref{Heating}), this maximum is clearly the result of this non-local transit effect. At 70 MHz many beam electrons hit the opposing collapsing sheath at a time, when the sheath voltage is still high enough to reflect them back into the bulk. This leads to an effective confinement of these energetic electrons in the bulk with  relatively low value of $\varepsilon_{\rm{e}}$ at 70 MHz (see fig. \ref{ModelTerms}), and a high plasma density (see fig. \ref{Density}). This is confirmed in figure \ref{MeanEEnergy}: the local maxima of the mean electron energy during the phases of sheath collapse do not reach the electrode at 70 MHz. At this frequency the maximum mean energy during sheath collapse, when an incoming beam is reflected, is stronger compared to the maximum during sheath expansion, when a new beam is generated. This is explained by two reasons: (i) The beam electrons hitting the collapse are more energetic compared to those hitting the expansion phase (see figures \ref{HighEnergy1} and \ref{HighEnergy2}) and (ii) the local density of the energetic beam electrons is increased by the fact that incoming and reflected beam electrons cross the same position at the same time within the RF period. The time it takes an energetic beam electron to be reflected from the collapsing sheath is about 2 - 3 ns. This corresponds to the time between an electron entering and leaving the sheath again after reversing its direction. At 70 MHz and higher frequencies this time is comparable to the duration of the sheath collapse. Consequently a beam electron entering the collapsing sheath shortly before the sheath collapse will leave the sheath again during sheath expansion. This is illustrated in figure \ref{MeanEEnergy} by the broad banana shaped maximum of the electron mean energy during sheath collapse at 70 MHz. 

Within the low density resonant heating mode at frequencies below 60 MHz two electron beams at each electrode per sheath expansion are generated. While most electrons of the first beam hit the opposing sheath at a phase of high local sheath voltage resulting in effective confinement, at 50 MHz most electrons of the second beam hit the opposing sheath during the sheath collapse. This is illustrated in figure \ref{MeanEEnergy} by the fact that the local maxima of the mean electron energy during sheath collapse reach the electrode surface at both electrodes. This results in a poor confinement of these energetic electrons and the observed increase of $\varepsilon_{\rm{e}}$ (see fig. \ref{ModelTerms}). This, in turn, results in a smaller plasma density.

Due to the lower plasma density at low driving frequencies the minimum and maximum sheath widths increase significantly such as shown in figure \ref{Sheath}. The sheath expansion velocity remains high so that the plasma remains in the resonant heating mode. Due to the higher plasma density and smaller sheath widths at high driving frequencies (caused by the modulated confinement quality of energetic beam electrons) the bulk length is larger at high driving frequencies. Thus, the spatial region of high electron conduction current is large and beam electrons dissipate more power via collisions before they reach the opposing electrodes. This results in less cooling per electron and in combination with the higher density itself in a higher value of $S_{{\rm{abs}}}$ at high driving frequencies. According to equation (\ref{Model}) this leads to a stronger density increase.

Due to the lower plasma density and larger maximum sheath widths at low driving frequencies the sheaths are more collisional at low compared to high driving frequencies. This explains the sudden increase of the ion velocity at the electrodes at 60 MHz shown in the right plot of figure \ref{ModelTerms}. The minimum sheath voltages during sheath collapse are comparable at 50 MHz, 60 MHz and 70 MHz such as shown in the right plot of figure \ref{Sheath}. 

A much stronger increase of $\bar{n}_{\rm{e}}$ compared to $n_{\rm{i,el}}$ is observed in figure \ref{Density}, since the bulk length and, thus, the region of high electron density is strongly reduced at low compared to high driving frequencies due to the modulated confinement quality of energetic beam electrons.

 Figure \ref{EEDF} shows the time averaged EEPF obtained from the simulation for the base case at different driving frequencies. The plot illustrates well that the electron density is significantly lower at 50 and 55 MHz compared to higher frequencies due to the different heating modes. It is also perceptible that the shape of the EEPF is significantly different at the different driving frequencies. For lower frequencies (50 and 55 MHz) the slope of the EEPF is smaller than for 60 MHz. Therefore, the mean electron energy is higher which in turn determines the collisional energy loss per electron-ion pair created, $\varepsilon_{c}$ (see figure \ref{ModelTerms}). If the frequency is increased from 60 MHz to 90 MHz (higher density mode) the shape of the EEPF also changes (see figure \ref{EEDF}, right).  
 Thus, the shape of the EEPF can be controlled by adjusting the frequency. This is important for potential applications, since it might allow to customize the plasma chemistry in more complex gas mixtures.

\begin{figure}[h!]
	\begin{center}
		\begin{tabular}{cc}
			\includegraphics[width=0.5\textwidth]{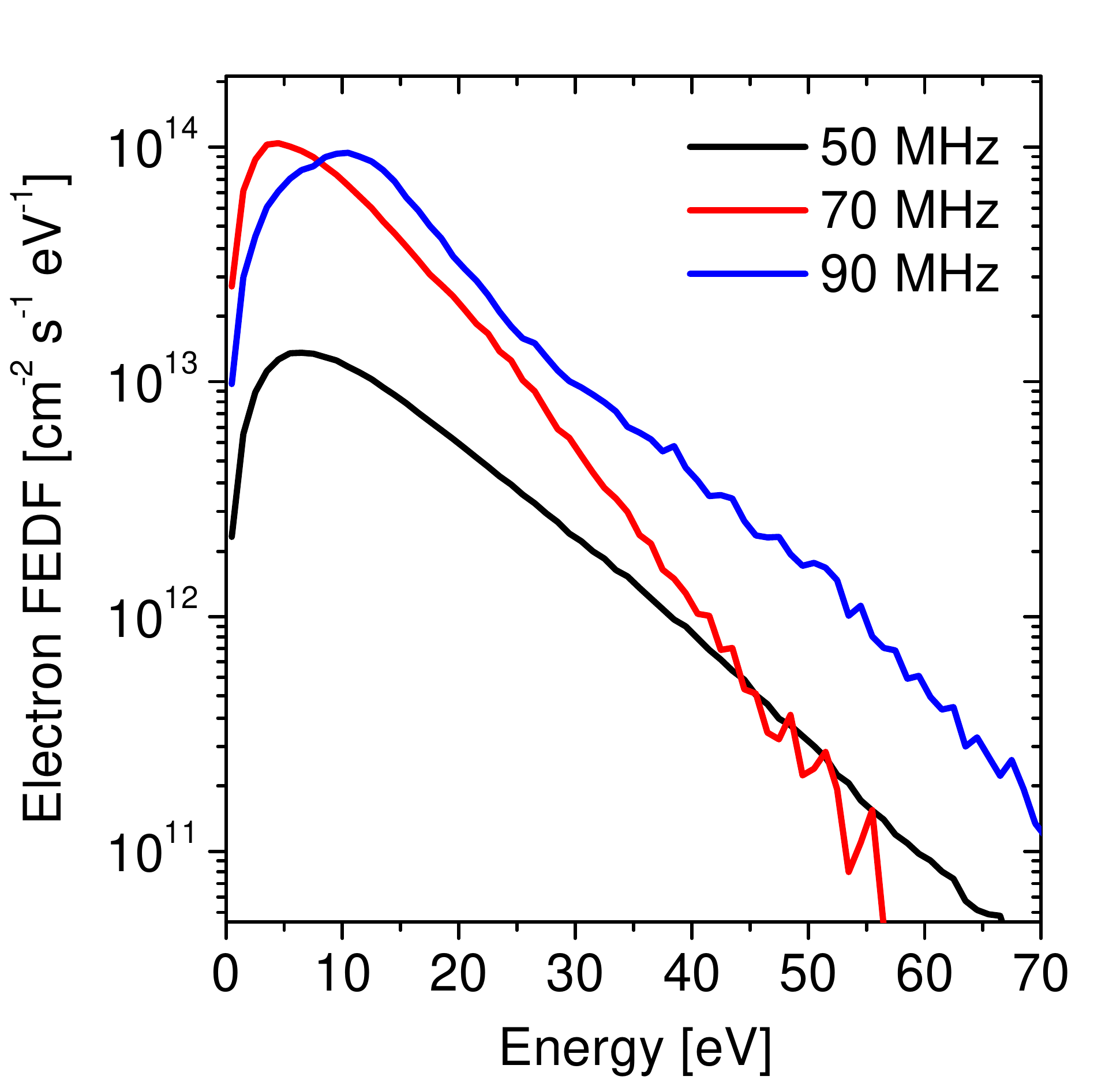}
			\\
		\end{tabular}
		\caption{Time averaged flux-energy distribution of the electrons lost at the electrodes for different driving frequencies obtained from the simulation. Conditions: $p = 1.3$ Pa, $d = 1.5$ cm, $\phi_{0} = 150$ V, $s = 1$, and $\gamma = 0$.}
		\label{FEDF}
	\end{center}
\end{figure}

Figure \ref{FEDF} shows the electron flux-energy distribution function (FEDF) at the electrodes. The plot shows that the fraction of highly energetic electrons lost at the electrodes is higher at 50 MHz compared to 70 MHz and 90 MHz due to the ineffective confinement of energetic beam electrons in the low density resonant heating mode characterized by the generation of multiple beams during one phase of sheath expansion.

Increasing the driving frequency leads to a shortening of the sheath collapse. This improves the confinement quality of beam electrons, since the time interval of electron loss to the electrodes is shortened. As more energetic electrons are reflected back into the bulk, the number of energetic electrons in the plasma increases. These beam electrons can be heated several times by multiple interactions with the boundary sheaths at both electrodes. Moreover, the shortening of the phase of sheath collapse at higher driving frequencies also leads to a higher electron heating on time average (see figures \ref{ModelTerms} and \ref{Heating}), since beam electrons hitting the opposing sheath during its collapse will leave the sheath again during the expansion time due to the finite time of 2 - 3 ns required to reverse the electron velocity in the sheath. The combination of these mechanisms causes the mean electron energy close to the electrodes to increase at very high driving frequencies such as shown in figure \ref{MeanEEnergy}. This, in turn, causes the energy lost per electron lost at the electrodes during the short phase of sheath collapse, $\varepsilon_{\rm{e}}$, to increase to values comparable to those found at lower driving frequencies around the step-like increase of the plasma density (see figure \ref{ModelTerms} and the right plot of figure \ref{EEDF}). However, at very high driving frequencies around 90 MHz, $S_{\rm{abs}}$ remains high and, therefore, the plasma density remains high, the sheath widths remain small and no heating mode transition is induced. 

\subsection{Dependence on electrode gap, pressure, and surface coefficients}

\begin{figure}[h!]
\begin{center}
\begin{tabular}{cc}
  \includegraphics[width=0.5\textwidth]{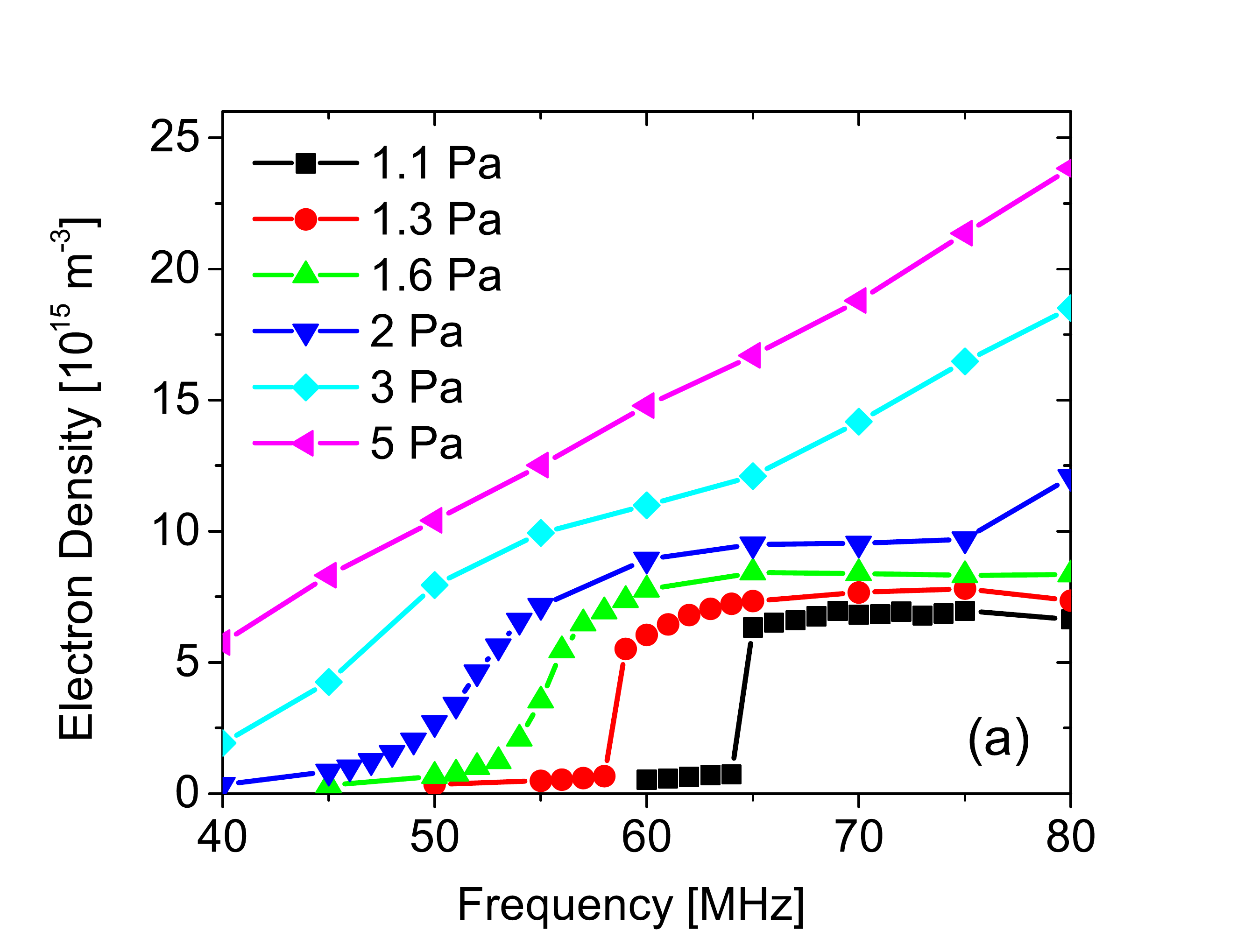}
 &
  \includegraphics[width=0.5\textwidth]{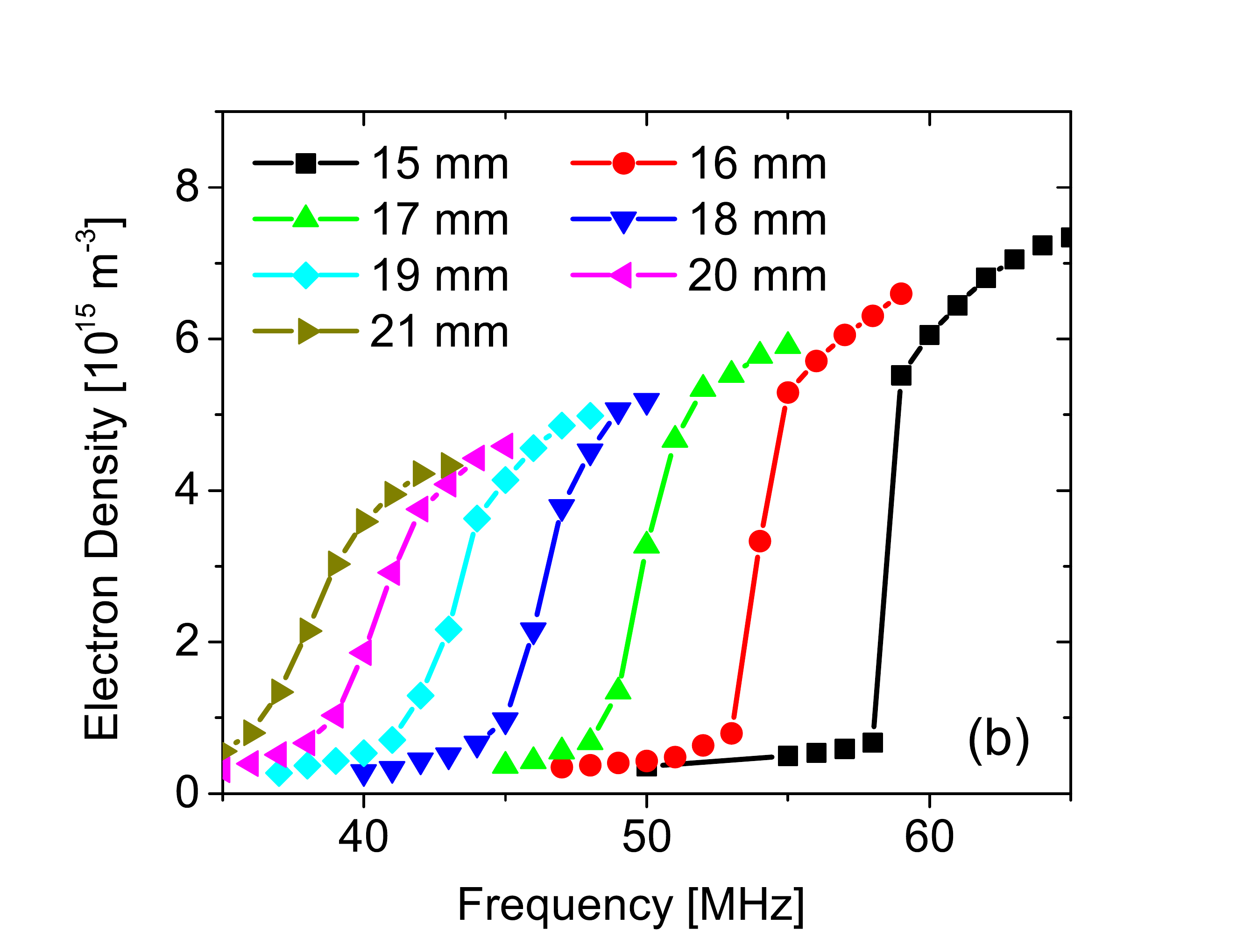}
 \\
\end{tabular}
\caption{Left: Space and time averaged electron density, $\bar{n}_{\rm{e}}$, as a function of the driving frequency for different neutral gas pressures at $d = 1.5$ cm. Right: $\bar{n}_{\rm{e}}$ as a function of the driving frequency for different electrode gaps at 1.3 Pa. $\phi_{0} = 150$ V, $s = 1$, and $\gamma = 0$ for all cases.}
\label{PressGapVar}
\end{center}
\end{figure}

Increasing the neutral gas pressure at otherwise constant input parameters in the simulation leads to an attenuation of the step-like increase of the plasma density such as shown in the left plot of figure \ref{PressGapVar}. This trend is caused by more collisional scattering of energetic beam electrons on their way from one electrode to the other and the higher plasma density at a given driving frequency. 
This leads to a thermalization and a broader energy distribution of beam electrons, when they arrive at the opposing electrode. Moreover, the sheath width is reduced at a given frequency due to the higher plasma density at higher pressures. Consequently the heating mode transition happens at lower driving frequencies.

	\begin{figure}[h!]
		\begin{center}
			\begin{tabular}{cc}
				\includegraphics[width=0.45\textwidth]{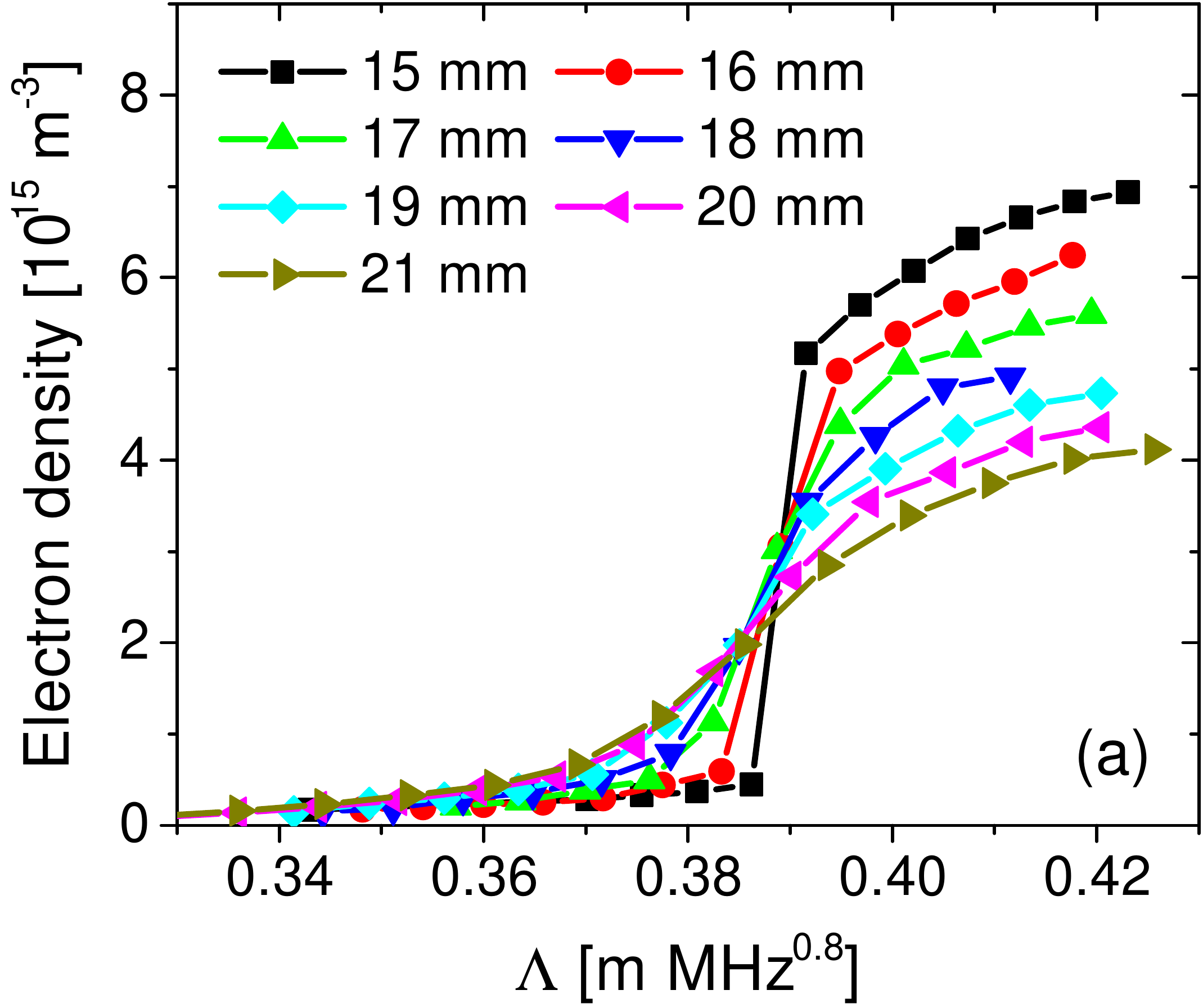}
				&
				\includegraphics[width=0.5\textwidth]{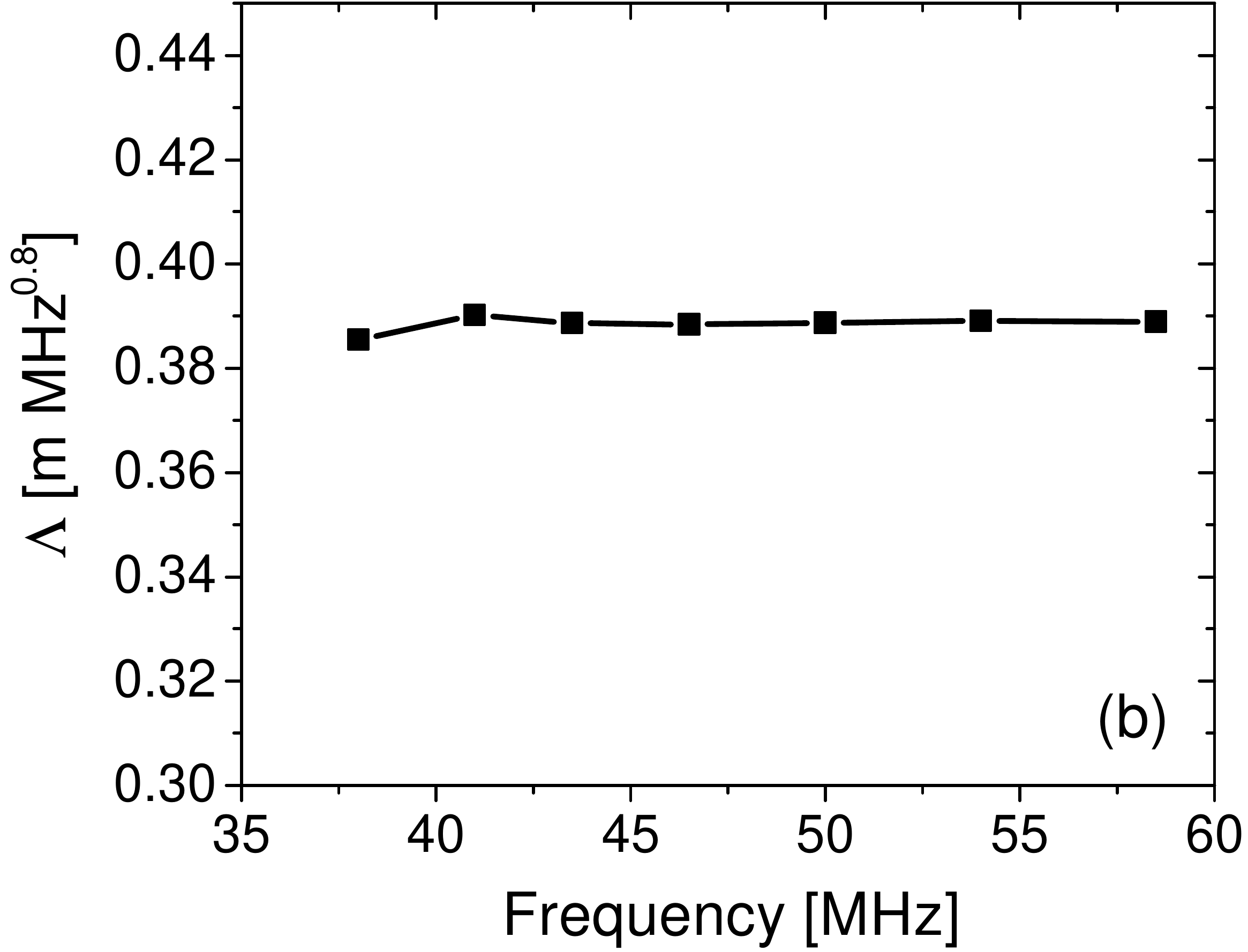}
				\\
			\end{tabular}
			\caption{Left: Average electron density as a function of $\Lambda$. Right: $\Lambda$ as a function of the transition frequency. Conditions: $p = 1.3$ Pa,  $\phi_{0} = 150$ V, $s = 1$, and $\gamma = 0$.}
			\label{gap_frequency}
		\end{center}
	\end{figure}

The right plot of figure \ref{PressGapVar} shows the averaged plasma density as a function of the driving frequency for different electrode gaps. A steeper increase of the density at higher driving frequencies is observed for short compared to large electrode gaps. The steeper increase of the density at small gaps is presumably caused by the reduced spread of the electron beam, when it reaches the opposing electrodes, i.e. its narrower energy distribution. This will cause a stronger modulation of the loss energy at the electrodes, if a beam hits the sheath collapse. For small gaps the density jump occurs at higher frequencies, because a beam electron propagating at a given velocity needs a shorter time to arrive at the opposing sheath due to the smaller electrode gap. In order to enable such beam electrons to hit the opposing sheath during its collapse and to cause the density jump in this way, the RF period must be shortened i.e. the driving frequency must be increased.
In order to find a criterion for the transition depending on pressure, frequency and gap size, a wider range of parameters must be investigated. Based on our results, we can define the following criterion for the step-like density increase at a fixed pressure of 1.3 Pa.
\begin{equation}
d\cdot f^{\alpha} \approx \Lambda ,
\end{equation}
	where $\alpha \approx 0.8$, and $\Lambda \approx 0.39$ m MHz$^{0.8}$.
	The left plot of figure \ref{gap_frequency} shows the average electron density as a function of $\Lambda$. The curves for different gap sizes cross each other at the transition at approximately $\Lambda = 0.39$ m MHz$^{0.8}$. The right plot presents $\Lambda$ versus the transition frequency, which was determined at the center of each transition from the left plot of figure \ref{PressGapVar}. The straight line confirms that the transition always occurs at $d \cdot f^{0.8} \approx 0.39$.
	This simple criterion must be generalized in future work in order to include dependencies on the pressure as well as the driving voltage.

\begin{figure}[h!]
\begin{center}
\begin{tabular}{cc}
  \includegraphics[width=0.5\textwidth]{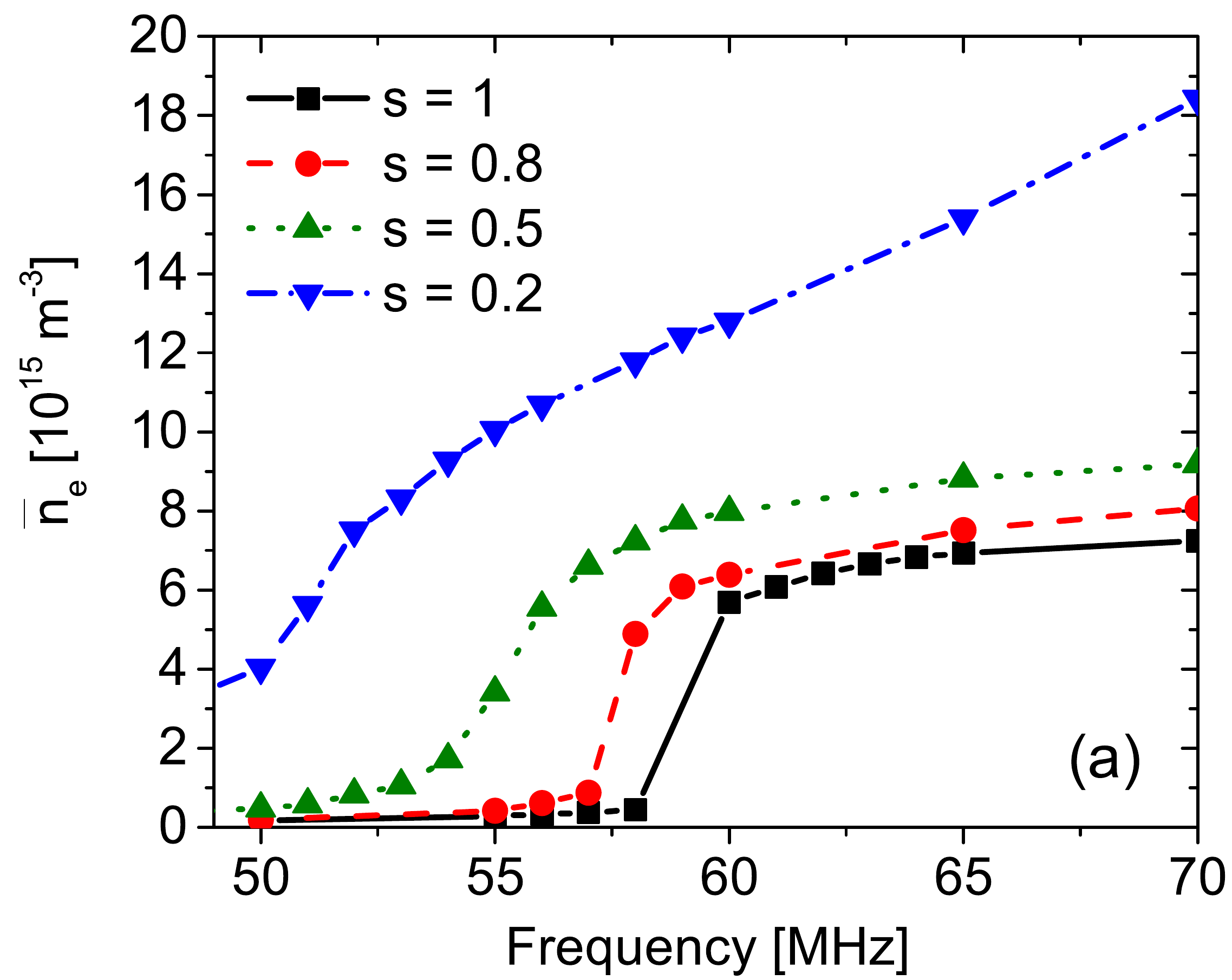}
 &
  \includegraphics[width=0.5\textwidth]{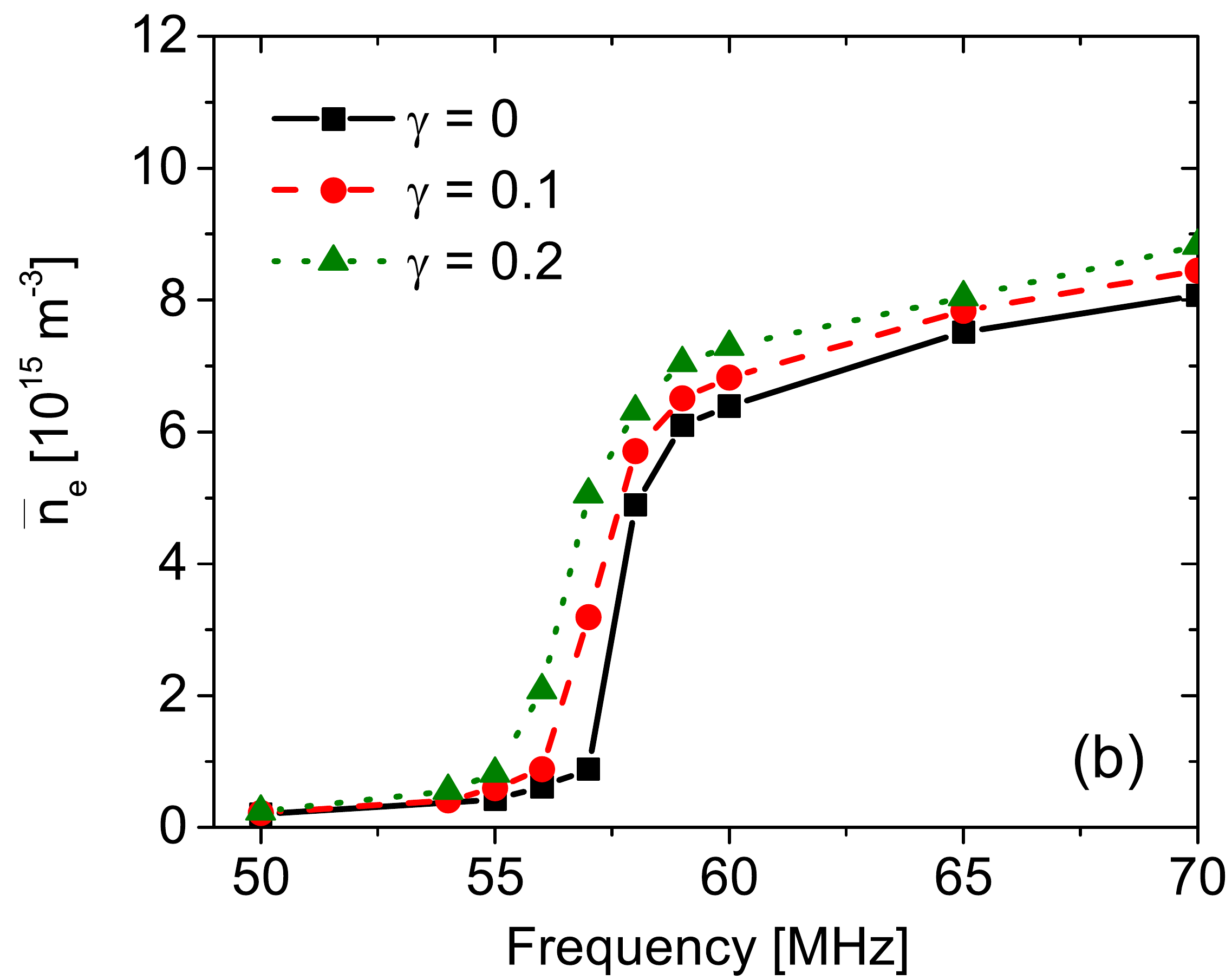}
 \\
\end{tabular}
\caption{Space and time averaged electron density as a function of the driving frequency for different electron sticking coefficients (left) and secondary electron emission coefficients (right) at the electrodes. Conditions: $p = 1.3$ Pa, $d = 1.5$ cm, $\phi_{0} = 150$ V. (right, $\gamma = 0$), (left, $s = 1$) .}
\label{Reflection}
\end{center}
\end{figure}

The left plot of figure \ref{Reflection} shows the averaged plasma density as a function of the driving frequency for different electron sticking coefficients at the electrodes. Typically a sticking coefficient of 0.8 is used in simulations of CCRF plasmas \cite{Schulze_Gamma}. Reducing the sticking coefficient leads to a smoother density increase, because more beam electrons are reflected after hitting the electrode during sheath collapse. The step-like increase of the density is observed at lower frequencies in case of a lower sticking coefficient, since more reflected electrons lead to a higher density at a given frequency. This results in a slower sheath expansion at a given frequency and leads to the heating mode transition causing the density jump to occur at lower driving frequencies, where the plasma density is lower. The deviations from a quadratic dependence of the plasma density on the driving frequency remain substantial even for very low sticking coefficients. The electron sticking coefficient depends on the electrode material and is unknown for many materials under plasma conditions. Comparing measured plasma densities as a function of the driving frequency to those obtained from the simulation under variation of $s$ might allow to determine the sticking coefficient for a given electrode material. 

The right plot of figure \ref{Reflection} shows the averaged plasma density as a function of the driving frequency for different $\gamma$-coefficients. Changing the secondary electron emission coefficient has only a weak effect on the plasma density under these low pressure collisionless conditions, since $\gamma$-electrons are not multiplied efficiently by collisions inside the sheaths. Increasing $\gamma$ leads to a weak increase of the plasma density and shifts the heating mode transition and, thus, the density jump towards lower driving frequencies for the same reasons as discussed for the variation of the electron sticking coefficient. 

\section{Conclusions}

In contrast to previous assumptions, the dependence of the plasma density on the driving frequency in low pressure single frequency capacitively coupled argon plasmas is not found to be quadratic. Instead, a strong step-like increase of the density at distinct driving frequencies is observed in kinetic PIC/MCC simulations. Based on an electron power balance model and a detailed analysis of the electron dynamics observed in the simulation, this increase is found to be related to a modulated confinement quality of energetic electron beams generated at each electrode during the phase of sheath expansion as a function of the driving frequency. This modulation results in a drastic change of the energy lost per electron lost at the electrodes and causes the observed density jump. It is related to an electron heating mode transition from the classical $\alpha$-mode at high driving frequencies and plasma densities into a low density resonant mode at lower driving frequencies. This mode transition is induced by a faster sheath expansion at lower frequencies and plasma densities on the timescale of the inverse local electron plasma frequency. Similarly to earlier results of Vender et al.\cite{VenderFieldRev} and Ku et al.\cite{Annaratone,Ku,Ku2} this causes high frequency oscillations of the electron heating rate. Consequently, the generation of two electron beams at each electrode per sheath expansion phase is observed in the simulations in contrast to one beam in the classical $\alpha$-mode observed at higher driving frequencies. At the low pressures investigated here these beams propagate through the plasma bulk almost without collisions and interact with the opposing RF boundary sheath. In the low density mode, the second beam generated at each electrode per RF period hits the opposing sheath during sheath collapse resulting in a poor confinement of those beam electrons and an increase of the energy lost by electrons lost at the electrodes in contrast to the high density mode, where a good confinement quality is observed. 

Due to the low plasma density the maximum sheath widths are large and the plasma bulk is small in the resonant heating mode. Thus, beam electrons launched at the sheath edge at one electrode cannot dissipate much energy via collisions while propagating through the bulk and are cooled immediately at the opposing  collapsing sheath. This results in a sudden decrease of the electron heating on time average in the low density mode and causes a stronger change of the plasma density. Due to their large widths the sheaths are more collisional in the low density regime and the ion velocity at the electrode decreases compared to the high density mode at higher driving frequencies.

At very high driving frequencies the confinement quality of energetic beam electrons is improved by shortening the duration of the sheath collapse at each electrode. Beam electrons entering the collapsing sheath are reflected and leave the sheath during sheath expansion effectively, resulting in less efficient electron cooling during sheath collapse and more electron heating on time average. Consequently, more energetic electrons are observed at high driving frequencies and the average energy of electrons absorbed at the electrodes during the short phase of sheath collapse is high. 

Increasing the neutral gas pressure is found to attenuate the density jump and to shift it towards lower driving frequencies due to collisional scattering of the beam electrons on their way to the opposing electrode and higher plasma densities. The latter effect causes the sheaths to expand more slowly at a given frequency and causes the heating mode transition to occur at lower driving frequencies, where the plasma density is lower and the sheaths expand faster. For similar reasons decreasing the electrode gap shifts the density jump towards higher frequencies and increases its height.

Decreasing the electron sticking coefficient results in a smoother density jump, since the modulation of the confinement quality of the energetic beam electrons is attenuated. Comparing density measurements as a function of the driving frequency with simulation results under variation of the sticking coefficient, therefore, potentially allows to determine the sticking coefficient for a given electrode material.

Changing the $\gamma$-coefficient has only a weak effect on the plasma density under the low pressure conditions investigated, since secondary electrons are not multiplied efficiently by collisions inside the sheaths. Besides the weak increase of the plasma density as a function of $\gamma$ a shift of the density jump towards lower driving frequencies is found, since the heating mode transition occurs at lower driving frequencies.

The observed step-like increase of the plasma density is expected to be highly relevant for the reactor design of low pressure CCRF plasmas used for plasma etching and RF sputtering, where engineers typically intend to maximize the plasma density and ion flux to optimize process rates, while keeping the driving frequencies low. 


\ack{This work has been supported by the German Research Foundation (DFG) within the frame of the Collaborative Research Center TRR 87 and by the Hungarian Scientific Research Fund through the grant OTKA K-105476.}



\end{document}